**How does international guidance for statistical practice align with the ASA Ethical Guidelines for Statistical Practice?**

**Rochelle E. Tractenberg[1]**

**Jennifer Park[2]**

1. Collaborative for Research on Outcomes and –Metrics; and Departments of Neurology; Biostatistics, Bioinformatics & Biomathematics; and Rehabilitation Medicine, Georgetown University, Washington, DC, USA

ORCID: 0000-0002-1121-2119

2. Committee on National Statistics, National Academies of Science, Engineering, and Medicine, Washington, DC USA

**Correspondence to:**

Rochelle E. Tractenberg
Georgetown University Neurology
Room 207, Building D
4000 Reservoir Rd., NW
Washington, DC, 20057 USA
**Email:** rochelle -dot- tractenberg -at -gmail -dot- com

Acknowledgement: There are no actual or potential conflicts of interest. Opinions expressed in this article are the authors' own.

Running Head: International Guidelines align with ASA Ethical Guidelines

**A. Objective**

    ***1. Principles matter***

Gillikin (2017) defines a "practice standard" as a document to "define the way the profession's body of knowledge is ethically translated into day-to-day activities" (Gillikin 2017, p. 1). Such documents fulfill three objectives: they 1) define the profession; 2) communicate uniform standards to stakeholders; and 3) reduce conflicts between personal and professional conduct (Gillikin, 2017 p. 2). Principles can inform program, staffing, and budgetary decisions across organizations and governments. However, there are many guidelines - this is due to different purposes that guidance writers may have, as well as to the fact that there are different audiences for the many guidance documents. The existence of diverse statements do not necessarily make it clear that there are commonalities; and while some statements are explicitly aspirational, professionals as well as the public need to know that ethically-trained practitioners follow accepted practice standards. The existence of ethical practice standards for statistics and data science may be helpful in supporting the implementation of more aspirational statements of ethical practice. Moreover, if and when bodies consider updating their guidance for ethical practice, identified areas of alignment among diverse guidance documents can facilitate the recognition of consensus.

    ***2. Understanding differences in national guidance provide opportunity***

When it comes to ethical statistical practice, the Ethical Guidelines for Statistical Practices maintained by the American Statistical Association (ASA, 2022) are a particularly useful tool for measuring and communicating professional expectations for ethical statistical and data science practice. (Park and Tractenberg, 2023)

- The guidelines are comprehensive in scope. They define "statistical practice" as activities including "designing the collection of, summarizing, processing, analyzing, interpreting, or presenting data and model or algorithm development and deployment."
- The guidelines also apply broadly to all members of the profession, including "all those who engage in statistical practice, regardless of job title, profession, level, or field of degree."
- The guidelines are sufficiently detailed to guide ethical conduct of professionals and their organizations. The guidelines specifically describe both the core responsibilities of the profession and ethical conduct across eight Principles. An appendix, describes the ethical obligations of organizations and institutions engaging in statistical practice.

Park and Tractenberg (2023) demonstrated that the ASA Ethical Guidelines align strongly with US legal, regulatory, and academic guidance for federal statistical and data science agencies. Nonetheless, areas of possible tension and guidance gaps were noted. These signaled opportunities to clarify language, provide examples for implementation, and collaborate for possible revision.

    ***3. Understanding differences in international guidance can provide mutual benefit***

The ASA Ethical Guidelines were first adopted in 1999. Consistent with best practice, the guidelines have been reviewed periodically and revised occasionally (most recently, 2022) to take into account changes in society and the profession, and to improve accessibility and therefore utility.

Roughly over this same time span, guidelines for official statistics were issued by several important collaborators in the international statistical system: the United Nations Fundamental Principles of Official Statistics (2014), the European Statistical System Committee European Code of Statistical Practice (2017), the Organization for Economic Cooperation and Development Good Statistical Practice (2014), and the International Statistical Institute Declaration on Professional Ethics (2023).

How does international guidance for statistical practice align with the ASA Ethical Guidelines? Taking into account organizational and constituency differences, how can understanding areas of convergence, possible tensions, and potential gaps inform practitioners and organizations relying on ethical statistical practice by all who engage with data? How does this understanding inform thinking about future revisions among the guidance documents considered?

### 4. Purpose of this paper

This paper applies the methodological approach described in Tractenberg (2023) and demonstrated in Park and Tractenberg (2023) to study alignment among international guidance for official statistics, and between these guidance documents and the ASA Ethical Guidelines for Statistical Practice functioning as an ethical practice standard (Tractenberg, 2022-A, 2022-B; after Gillikin 2017). In the spirit of exchanging experiences and lessons learned, we discuss how our findings could inform closer examination, clarification, and, if beneficial, possible revision of guidance in the future.

## B. Data

### 1. American Statistical Association Ethical Guidelines for Statistical Practice (ASA EG)

The American Statistical Association is one of the oldest continuously operating professional organizations for statistical practice in the world. It was founded in the US in (1839) and has had international membership ever since. It was established to promote the practice and profession of statistics - in the United States and worldwide.

The ASA Ethical Guidelines for Statistical Practice represent the ethical practice standard for statistics and data science (Tractenberg, 2022-A, 2022-B). These ethical practice standards were originally adopted by the ASA in 1995 (Hogan & Steffey) and have been periodically updated since 2016. The most recent update (2022) includes 60 items organized under eight principles and a 12-item Appendix specific for organizations and institutions. See Box 1. The next anticipated revision will occur in 2027.

## Box 1. ASA Ethical Guidelines for Statistical Practice, 2022

| |
|---|
| A. Professional Integrity & Accountability (12) |
| B. Integrity of data and methods (7) |
| C. Responsibilities to Stakeholders (8) |
| D. Responsibilities to research subjects, data subjects, or those directly affected by statistical practices, Data Subjects, or those directly affected by statistical practices (11) |
| E. Responsibilities to members of multidisciplinary teams (4) |
| F. Responsibilities to Fellow Statistical Practitioners and the Profession (5) |
| G. Responsibilities of Leaders, Supervisors, and Mentors in Statistical Practice (5) |
| H. Responsibilities regarding potential misconduct (8) |
| APPENDIX: Responsibilities of organizations/institutions (12) |

These Guidelines are comprehensive, and actionable - not aspirational. Studies conducted by the authors have shown alignment of the ASA Ethical Guidelines with ethical guidance from the Royal Statistical Society (2014) (Tractenberg, 2020), the Association for Computing Machinery (2018) (Tractenberg, 2022-A), US law [Evidence Act of 2018) and policy (2018, 2021). (See Park & Tractenberg 2023). Given this strong alignment,[1] the authors recommend the ASA Ethical Guidelines as a useful, actionable tool for communicating and monitoring ethical statistical practice to practitioners using statistics within a range of disciplines, and for audiences in the US.

Tractenberg 2022-A, 2022-B refers to the ASA Ethical Guidelines as an "ethical practice standard for statistics and data science". An appendix of 12 items describe obligations of organizations and institutions that utilize, or contract for, statistical practices to support and prioritize ethical statistical practices. This subset of the Ethical Guidelines[2] are unique among ethical practice guidelines for statistics, data science, and data ethics. They formally recognize that individual practitioners' roles and responsibilities can change across a career -and that their ethical obligations change as well. By specifying ethical obligations of leaders, mentors, and supervisors as well as of organizations and institutions employing statistical practices and practitioners, the ASA EGSP are particularly relevant for this study where international guidelines for ethical statistical practices within and across agencies are considered.

### 2.  United Nations Fundamental Principles of Official Statistics (UN FPOS)

The need for a set of international principles governing official statistics became apparent at the end of the 1980s when countries in Central Europe began to change from centrally planned economies to market-oriented democracies. It was essential to ensure that national statistical systems in such countries would be able to produce appropriate and reliable data that adhered to certain professional and scientific standards. Towards this end, the Conference of European Statisticians developed and adopted the Fundamental Principles of Official Statistics in 1991 (CES/702), which were subsequently adopted in 1992 at the ministerial level by ECE as decision C(47). Following an international consultation process, the United Nations Statistical Commission adopted these principles as the United Nations Fundamental Principles of Official Statistics in 1994. The Economic and Social Council adopted the draft resolution on the Fundamental Principles of Official Statistics in its resolution 2013/21 of 2013. The highest authoritative body in the UN, the General Assembly, endorsed the FPOS in its resolution 68/261 of 2014.[3] See Box 2.

As principles, the FPOS notably have remained unchanged (with the exception of the preamble) since adoption. To provide additional reference and assist with transparency, a Friends of the Chair group reporting to the Chair of the UN Statistics Commission engaged in a global consultation for implementation guidance. A 2015 report reflected inputs from 40 countries. Periodic global surveys have described country practices in implementation.

---

[1] Five exceptions to alignment were found (one aspect of the Royal Statistical Society Code; two computing-specific, and one membership-specific aspects of the Association of Computing Machinery Code; and one element of the International Statistics Institute Statement).

[2] Although organizations and institutions (employers) are discussed in the ASA Ethical Guidelines Appendix, they are grouped together into a single column with ASA Principle G in our alignment tables because of the potential overlap of employers or leaders who are not statistical practitioners (guidance in Appendix) and leaders, supervisors, and mentors who are statistical practitioners (guidance given in Principle G).

[3] https://unstats.un.org/unsd/dnss/gp/fundprinciples.aspx.

**Box 2. UN Fundamental Principles of Official Statistics, 2014**

| |
|---|
| 1. Official statistics that meet the test of practical utility are to be compiled and made available publicly. |
| 2. Statistical agencies choose methods and procedures for the collection, processing, storage and presentation of statistical data based on strictly professional considerations, including scientific principles and professional ethics. |
| 3. Statistical agencies are to present information according to scientific standards on the sources, methods and procedures of the statistics. |
| 4. The statistical agencies are entitled to comment on erroneous interpretation and misuse of statistics. |
| 5. Statistical agencies choose data sources with regard to quality, timeliness, costs and the burden on respondents. |
| 6. Confidential data collection used exclusively for statistical purposes. |
| 7. Public laws, regulations and measures under which the statistical systems operate |
| 8. Coordination among statistical agencies within countries |
| 9. The consistent use by statistical agencies in each country of international concepts, classifications and methods |
| 10. Statistical agencies cooperate bilaterally and multilaterally |

### 3. European Statistical System Committee European Statistics Code of Practice (ESSC CoP)

The European Statistical System Committee is a partnership in which Eurostat and the national statistical authorities of each European Union member states and European Free Trade Association states cooperate. The joint mission of these entities is to provide independent high quality statistical information at European, national and regional levels. Each commits themselves to adhere to the European Code of Practice.

The ESSC is the steward of the European Statistics Code of Practice,[4] a self-regulatory quality assurance instrument based on 16 principles spanning the institutional environment, statistical processes and statistical outputs. A set of indicators of best practices and standards for each of the principles provides guidance for reviewing implementation, with the aim of improving adherence and transparency.

An independent Advisory Board, the European Statistical Governance Advisory Board, analyzes the implementation of the Code of Practice by the European Union Statistical Authority (Eurostat) and the European Statistical System as a whole every year; and advises on the implementation of the Code of Practice as well as on its possible updates.

First adopted in 2005, the 2017 edition reflects innovations in the development, production and dissemination of official statistics in the European Statistical System, such as emerging new data sources, use of new technologies, modernisation of the legal framework, and the results of the peer reviews.

---

[4] European Statistics Code of Practice 2017 KS-02-18-142-EN-N.pdf.

**Box 3. European Statistical System Committee European Statistics Code of Practice, 2017**

| Institutional environment |
|---|
| 1. Professional Independence |
| 1bis. Coordination and cooperation |
| 2. Mandate for Data Collection and Access to Data |
| 3. Adequacy of Resources |
| 4. Commitment to Quality |
| 5. Statistical Confidentiality and Data Protection |
| Statistical Processes |
| 6. Impartiality and Objectivity |
| 7. Sound Methodology |
| 8. Appropriate Statistical Procedures |
| 9. Non-excessive Burden on Respondents |
| 10. Cost Effectiveness |
| Statistical Output |
| 11. Relevance |
| 12. Accuracy and Reliability |
| 13. Timeliness and Punctuality |
| 14. Coherence and Comparability |
| 15. Accessibility and Clarity |

### 4. *Organization for Economic Cooperation and Development Good Statistical Practice (OECD GSP)*

The OECD is a private, non-governmental organization which convenes governments to address the economic, social and environmental challenges of globalization. The Organisation provides a setting where governments[5] can exchange policy experiences, seek answers to common problems, identify good practice and work to coordinate international policies. New members are engaged through a review and ascension process during which petitioning countries exchange experiences and descriptions about their national infrastructure, policies, and priorities.

In Article 3 of the OECD Convention, members agreed to "furnish the Organisation with the information necessary for the accomplishment of its tasks." The quality of statistics is fundamental for the quality of evidence-based analytical work of the Organisation and for the quality of statistical publications and databases produced by the OECD.

Adopted by the OECD Council in 2015 on the proposal of the OECD Committee on Statistics and Statistical Policy (CSSP), the Recommendation provides a detailed blueprint for a sound and credible national statistical system. Each of the 12 specific recommendations is supplemented by a set of indicative good practices, which together provide a framework for examining the implementation of the Recommendation in practice. The OECD GSP Recommendations are presented as a series of yes/no checkboxes oriented to the concerns of national statistical offices, rather than offering guidance for ethical statistical practice at the individual practitioner level.

---

[5] The OECD Member countries are: Australia, Austria, Belgium, Canada, Chile, Colombia, Costa Rica, the Czech Republic, Denmark, Estonia, Finland, France, Germany, Greece, Hungary, Iceland, Ireland, Israel, Italy, Japan, Korea, Latvia, Lithuania, Luxembourg, Mexico, the Netherlands, New Zealand, Norway, Poland, Portugal, the Slovak Republic, Slovenia, Spain, Sweden, Switzerland, Türkiye, the United Kingdom and the United States.

The Recommendation also provides a benchmark against which the national statistical system of countries can be assessed, and constitutes a tool for self-assessment of non-Members which can facilitate the identification of needed improvements in their national statistical system. Individual assessments of national statistical systems could take several forms: a simple self-assessment, an evidence-based self-assessment and a peer review by the CSSP.

Periodic reports[6] to Council on the implementation of the Recommendation are based on information gathered through a number of tools, including a survey among Adherents to the Recommendation, a peer review of national statistical systems, and self-assessment questionnaires provided by Adherents. Reviews describe good practices and common challenges. The next report is anticipated in 2025.

### Box 4. OECD Good Statistical Practice, 2014

| |
|---|
| 1. A clear legal and institutional framework. |
| 2. Professional independence. |
| 3. Adequacy of resources. |
| 4. Protection of privacy. |
| 5. The right to access administrative sources. |
| 6. Impartiality, objectivity and transparency. |
| 7. Sound methodology and professional standards. |
| 8. The quality of statistical outputs and processes. |
| 9. User-friendly access and dissemination of data and metadata, and a commitment to respond to major misinterpretations of data by users. |
| 10. Coordination of statistical activities. |
| 11. International cooperation. |
| 12. Exploring new and alternative data sources and methods |

5. ***International Statistics Institute Declaration on Professional Ethics (ISI DPE)***

The International Statistical Institute is a non-profit, non-government organization established in 1885 with individual and institutional members in over 150 countries. Its main objective is to promote the understanding, development, and good practice of statistics worldwide to exchange advances in statistical knowledge, best practices, and by creating opportunities to network.

The ISI Declaration on Professional Ethics was first issued in 1985, and subsequently revised in 2010 and most recently, 2023.

The aim of the Declaration is "to enable the statistician's individual ethical judgments and decisions to be informed by shared values and experience..." It "seeks to document widely held principles of the statistical profession and to identify the factors

---

[6] OECD, Recommendation of the Council on Good Statistical Practice, OECD/LEGAL/0417.

that obstruct their implementation" and offers an aspirational framework to guide this work.

Rather than implementation guidelines per se, the ISI provides a description of roles and strategies with regard to professional ethics. These can be summarized as ways to communicate support to professional statisticians, note concern to national authorities and international agencies, and to collaborate on joint communications with other professional societies where possible to amplify messages.

**Box 5. International StatistIcs Institute Declaration on Professional Ethics, 2023**

| Shared Professional Values |
| --- |
| 1. Respect |
| 2. Professionalism |
| 3. Truthfulness and Integrity |
| Ethical Principles |
| 1. Pursuing Objectivity |
| 2. Clarifying Obligations and Roles |
| 3. Assessing Alternatives Impartially |
| 4. Conflicting Interests |
| 5. Avoiding Preempted (pre-determined) Outcomes |
| 6. Guarding Privileged Information |
| 7. Exhibiting Professional Competence |
| 8. Maintaining Confidence in Statistics |
| 9. Exposing and Reviewing Methods and Findings |
| 10. Communicating Ethical Principles |
| 11.Bearing Responsibility for the Integrity of the Discipline |
| 12. Protecting the Interests of Subjects (human only, is implied) |

## C. Methods

### 1. *Degrees of freedom analysis*

Examining qualitative data, such as professional guidance, in a systematic, rigorous, but easily communicated manner is challenging. The Degrees of Freedom Analysis (DoFA, Campbell 1975; Tractenberg 2023) approach is a structured approach designed to facilitate the analysis of qualitative information into a coherent, interpretable, analysis framework. It is a form of qualitative content analysis where the investigator searches for patterns and themes in narrative (Nieswiadomy & Bailey, 2018). Originally intended for theory building (Campbell 1975), this method was adapted for decision making by Tractenberg (2019) and was applied to the analysis of alignment of ethical practice standards across diverse disciplines in Rios et al. (2019) and Tractenberg (2020). As outlined in Tractenberg (2019, 2023) and demonstrated in Tractenberg (2020, 2022-A, 2022-B), the matrix of the evidence or information to be studied is assembled into a two-dimensional table. In this paper we explore the thematic alignment between ethical guidelines as outlined in the previous section. First we examine alignment of each international guidelines, principles, and responsibilities for "good" or ethical federal statistical practices with the ASA Guideline Principles (ASA 2022). The Degrees of Freedom Analysis method (Tractenberg, 2023), used in Park & Tractenberg 2023 and extensively in Tractenberg (2022-A, 2022-B) was used for these analyses of alignment.

### 2. Coding schema

Recall (as described in the section above), the guidance documents examined for this analysis were developed for different purposes and different audiences. Thus, observations regarding alignment, tensions, and potential conflict should take these differences into account in coding schemes and interpretation of results. Accordingly, we adjusted our coding schema to capture these nuances.

For this analysis, we determined the coding scheme to describe alignment, tension, or potential conflict across each pair of guidelines examined *a priori;* the results are summarized in two-dimensional tables. To generate a consistent and interpretable "signal", all coding is in terms of the ASA EG. In other words, coding identifies specific elements of the ASA EG that align with (i.e., agree with, support) each of the elements from the other examined guidance document. Areas of tension, and potential conflict were also identified. The next section provides more detail regarding specific coding schema.

Box 6 provides a key to our coding schema. Definitions of terms and examples from our observations of guideline pairs are provided alongside exemplar detailed codes and summary codes. For example, a particular observation in the examination of a guidance document pair could indicate a scenario in which similar ideas, expressed as goals, products, or behavior, will be supportive of one another across guidance documents. These "alignments" therefore may be literal or similar matches (see (a) and (b), below). If a pattern of support is found, the relevant row or column is shaded in green (see (f) below).

In other cases, a particular observation could indicate a scenario in which certain ideas (again, expressed as goals, products, or behavior) across guidance documents are opposed (see (c) below). These "conflicts" may appear alone, or co-exist with areas of alignment so the resulting observation describes "tension" (see (d) below). If a pattern of tension (conflict and alignment) is found, the relevant row or column is shaded in yellow (see (g) below).

There are also potential scenarios of apparent "guidance gaps" within a particular guidance document that are represented in the other guidance document for a given pair. For a given instance, this is shown by a blank cell (see (e) below). Patterns of this scenario are shown by shading the relevant row or column in gray (see (h) below).

In sum, our analysis of alignment between the ASA Guidelines and the statements relating to official statistics reflect whether: 1) the ideas expressed are similar and/or supportive of one another; 2) the behaviors in the ASA Guidelines, when observed, will lead to/demonstrate the OECD, UN, or EU guidance principle; and/or 3) the practitioner who seeks to follow the ASA Ethical Guidelines may be hindered by one or more of the OECD, UN, EU, or ISI guidance principles or elements. Because it represents the ethical practice standard governing statistical practice ("defined as, "statistical practice" includes activities such as: designing the collection of, summarizing, processing, analyzing, interpreting, or presenting, data; as well as model or algorithm development and deployment." ASA 2022), the content analysis result is the identification of specific ASA Ethical Guideline element(s) that match the row/element of the other document.

We also include alignment analyses across the international guidance documents (FPOS, GSP, CoP, DPE), to document the extent to which these are aligned with each other, and using the same content analysis approach to determine if following any of these guidance documents might lead to tension or if gaps might be observed to exist.

While it is important to determine if individuals who follow the ASA Ethical Guidelines for Statistical Practice might encounter conflict with the other guidance documents, it is also important to recognize that when practitioners are collaborating with others who might be following different international guidance (e.g., UN and Eurostat) might need to consider how best to reconcile any tension (or gaps) that arise between their respective guidance documents.

### 3. Code assignment

As a first step, the first author examined each guidance pair and applied detailed codes. Once the coding was completed, the second author evaluated the results to discuss the contents or coding of all cells in that table. If the two authors disagreed on any match (schema described below), further discussion determined if consensus could be reached, or if that result should be identified as an area where further clarity would improve the interpretation and use of guidance documents. Detailed coded tables are presented in the Appendix.

Subsequently, the detailed coded tables were summarized into themes illustrated using a simplified coding strategy to improve communication of results; these summary tables are presented in the body of the report. (See below.) Note, although it is possible that different coders may generate somewhat different detailed codes than those prepared by the authors,[7] we are confident that the pattern of themes summarized in this paper would be consistent, assuming familiarity with the guidelines examined.

**Box 6. Key to alignment model coding schema**

| | Scenario | Example observation | Example detail code | Example summary code |
|---|---|---|---|---|
| a | Literal or conceptual match | If there is a thematic or exact match to the particular guideline, the particular ASA EG element is indicated | D[8] | ✔ [check] |
| b | Similar match | If the alignment with a particular guideline is abstract, or held under only specific circumstances, then the particular ASA EG element is indicated in parenthesis | (D)[9] | (✔)([check]) |

---

[7] See Tractenberg and Gordon (2017) for a fuller discussion of DoFA and its reliability and validity across content types/narratives.

[8] Example: ASA Ethical Guidelines A2: "(the ethical statistical practitioner) Uses methodology and data that are valid, relevant, and appropriate, without favoritism or prejudice, and in a manner intended to produce valid, interpretable, and reproducible results." (A2) and UN FPOS element 1: "official statistics that meet the test of practical utility are to be compiled and made available publicly."

[9] Example: ASA Ethical Guidelines A1: "(the ethical statistical practitioner Takes responsibility for evaluating potential tasks, assessing whether they have (or can attain) sufficient competence to execute each task, and that the work and timeline are feasible. Does not solicit or deliver work for which they are not qualified, or that they would not be willing to have peer reviewed", and OECD GSP 3 "Adequacy of resources."

| | Scenario | Example observation | Example detail code | Example summary code |
|---|---|---|---|---|
| c | Opposition | If a particular guideline could cause conflict, and attention is needed for balanced implementation of guidance, the ASA EG element appears in red font with a ~ ("not") | ~D10[10] | ~ [not] |
| d | Tension | If both alignment and potential conflict coexist for a particular guideline, then both ASA EG aligned and opposed elements are indicated | A2, ~A3, ~A4[11] | ✔[check], ~ [not] |
| e | Guidance gap | If there is neither exact nor thematic alignment (or nonalignment) | [blank][12] | [blank] |
| f | Complete alignment | Patterns of complete alignment across that row (or column) | D[13] | ✔[check], (✔) ([check]) |
| g | ASA EG Principle alignment | All ASA EG Principle elements (e.g., all 11 elements of A) are supportive of the corresponding guideline element (row- so the cell is dark green). | D[14] | ✔ [check], (✔) ([check]) |

## D.  Results

Summary tables are presented and described here. See the Appendix for detail tables.

### 1.  Alignment of ASA Ethical Guidelines and UN Fundamental Principles of Official Statistics

*Target audiences:* See Table 1. Recall that the UN FPOS were developed to communicate high-level expectations of leadership across nearly 200 national statistical offices to the broad international community. Accordingly, these principles are intended to pertain to a select portion of the professional statistician and data science community. The ASA EGs are intended for use by any statistics practitioner: "In these Guidelines, "statistical practice" includes activities such as: designing the collection of, summarizing, processing, analyzing, interpreting, or presenting, data; as well as model or algorithm development and deployment. Throughout these Guidelines, the term "statistical practitioner" includes all those who engage in statistical practice, regardless of job title, profession, level, or field of degree. The Guidelines are intended for individuals, but these principles are also relevant to organizations that engage in statistical practice."

---

[10] Example: ASA Ethical Guidelines A2, "(the ethical statistical practitioner) Uses methodology and data that are valid, relevant, and appropriate, without favoritism or prejudice, and in a manner intended to produce valid, interpretable, and reproducible results." may create tension with UNP 5 #1, "Statistical agencies choose data sources with regard to quality, timeliness, costs and the burden on respondents."

[11] Example: Eurostat elements 14 and 15 are supported by most ASA Ethical Guideline Principles, but in some cases tensions are also identified where following the Eurostat guidance could create conflict with ASA EGSP elements.

[12] Example: ASA Ethical Guidelines responsibilities relating to other statistical practitioners and the profession (Principle F) offers no support or guidance for how to ethically accomplish OECD 12, "Exploring new and alternative data sources and methods."

[13] Example: every ASA Ethical Guidelines Principle is supportive of Eurostat Institutional Environment element #4, "Commitment to Quality."

[14] Example: Every element in ASA Principle E (Responsibilities to members of multidisciplinary teams) is supportive of ISI Ethical Principle 1, "Pursuing Objectivity."

*Areas of strong alignment:* There was complete alignment of UN FPOS Principles 2 (*collection methods)* and 3 (*presentation standard)* with ASA EGs. That is, both of these Principles are aligned with all eight ASA EG principles; this pattern is denoted by green shading. Further, the particular practices and behaviors articulated by ASA EG Principles directly support achievement of all ten UN FPOS, notably, ASA EG C (*stakeholders*), and D (*subjects*). Nine of ten UN FPOS are supported by Principles A (*accountability*), B (*Integrity of data and methods*), E (*responsibilities to other disciplines*) and G (*responsibilities for leaders). (*See Appendix, Detail Table 1.)

*Areas of potential tension:* There was only one area of potential tension noted: when considering data on the basis of timeliness, cost, and respondent burden (UN FPOS 5 (*data selection*)), practitioners should also carefully consider their obligations under ASA EG Principle A (*accountability*), which describes professional expectations and integrity. Specifically, the ASA EG element A. 2 specifies that the ethical statistical practitioner "uses methodology and data that are valid, relevant, and appropriate, without favoritism or prejudice, and in a manner intended to produce valid, interpretable, and reproducible results. " This may cause tension for the practitioner who must balance the validity and reproducibility of their work against data that is "timely" or creates a low(er) burden on respondents.

| UN Fundamental Principles of Official Statistics / ASA Ethical Guidelines | A Accountability | B Integrity | C Stakeholder | D Subjects | E Other Disciplines | F Other Statisticians | G Leadership | H Misconduct |
|---|---|---|---|---|---|---|---|---|
| 1 Make data publicly available | ✓ | ✓ | ✓ | ✓ | ✓ | | ✓ | ✓ |
| 2 Collect using scientific methods | ✓ | ✓ | ✓ | ✓ | ✓ | ✓ | ✓ | ✓ |
| 3 Present using professional standards | ✓ | ✓ | ✓ | ✓ | ✓ | ✓ | ✓ | ✓ |
| 4 Comment on misuse | ✓ | ✓ | ✓ | ✓ | ✓ | | ✓ | ✓ |
| 5. Choose data considering quality, timeliness, cost, and respondent burden | (~) | ✓ | ✓ | ✓ | ✓ | | ✓ | ✓ |
| 6. Use confidential data only for statistical purposes | | | ✓ | ✓ | | | | |
| 7. Operate the statistical system within a legal framework | ✓ | ✓ | ✓ | ✓ | ✓ | | ✓ | ✓ |
| 8. Coordinate across national statistical agencies | ✓ | ✓ | ✓ | ✓ | ✓ | ✓ | ✓ | |
| 9. Use international statistical classifications consistently | ✓ | ✓ | ✓ | ✓ | ✓ | ✓ | ✓ | |
| 10. Cooperate across national statistical systems | ✓ | ✓ | ✓ | ✓ | ✓ | ✓ | ✓ | |

Table 1. Summary of Themes Across ASA EG and UN FPOS

*Guidance gap:* The UN FPOS Principle requiring that confidential data be used only for statistical purposes (Principle 6) is highly specific for official statistics, and is not a part of the ASA EG. The General Data Protection Regulation (GDPR; European Union 2018) Article 5.1(b) offers both the same explicit requirement and *also* allows for re-use of data in certain circumstances: "(Personal data shall be) collected for specified, explicit and legitimate purposes and not further processed in a manner that is incompatible with those purposes; further processing for archiving purposes in the public interest, scientific or historical research purposes or statistical purposes shall, in accordance with Article 89(1), not be considered to be incompatible with the initial purposes ('purpose limitation')". The GDPR "purpose limitation" is consistent with ASA EG Principles relating to stakeholders and data donors, and those affected by statistical practices (C and D); together these may address situations where a guidance gap is perceived between UN and ASA guidance on data use/limitation of use. Note also that the UN FPOS 1 requirement to "make data publicly available" necessarily limits the practitioner's ability to follow UN FPOS 6 (use confidential data solely for statistical purposes). Both the ASA EGs and GDPR are more specific about how to share - and use shared - data publicly.

*Discussion:* There was strong alignment of the ASA EG and the UN FPOS, with few gaps, and minimal potential for tension. Although these guidelines were developed for different audiences and with somewhat different objectives, the guidelines contained in each appear to be mutually reinforcing. We note that the UN FPOS explicitly calls for adherence to professional standards, such as the ASA EG.

## 2. Alignment of ASA Ethical Guidelines and ESSC European Statistical Code of Practice

| ESSC European Code of Statistical Practice / ASA Ethical Guidelines | A Accountability | B Integrity | C Stakeholder | D Subjects | E Other Disciplines | F Other Statisticians | G Leadership | H Misconduct |
|---|---|---|---|---|---|---|---|---|
| **Institutional Environment** | | | | | | | | |
| 1 Professional independence | ✓ | ✓ | ✓ | ✓ | ✓ | ✓ | ✓ | ✓ |
| 1bis Coordination and cooperation | ✓ | ✓ | ✓ | ✓ | ✓, ~ | ✓ | ✓ | ✓ |
| 2 Mandate for data collection and access to data | ✓, ~ | ✓ | ✓ | ✓ | ✓ | | | |
| 3 Adequacy of resources | (✓) | | | | | | ✓ | |
| 4 Commitment to quality | ✓ | ✓ | ✓ | ✓ | ✓ | ✓ | ✓ | ✓ |
| 5. Statistical confidentiality and data protection | ✓ | ✓ | ✓ | ✓ | ✓ | | | |
| **Statistical Processes** | | | | | | | | |
| 6. Impartiality and objectivity | ✓ | ✓ | ✓ | | ✓ | ✓ | ✓ | |
| 7. Sound methodology | ✓ | ✓ | ✓ | ✓ | ✓ | | ✓ | ✓ |
| 8. Appropriate statistical procedures | ✓ | ✓ | ✓ | ✓ | ✓ | | ✓ | ✓ |
| 9. No undue respondent burden | | | | (✓) | | | | |
| 10. Cost effectiveness | | | | (✓) | | | (✓) | |
| **Statistical Output** | | | | | | | | |
| 11. Relevance | ✓, ~ | ✓, ~ | ✓, ~ | (~) | ✓ | | ✓ | |
| 12. Accuracy and reliability | ✓ | ✓ | ✓ | | ✓ | | ✓ | |
| 13. Timeliness and punctuality | ~ | | | ~ | ~ | | ✓, ~ | |
| 14. Coherence and comparability | ✓, ~ | ✓, ~ | ✓ | (~) | ~ | ✓ | ✓, ~ | (✓) |
| 15. Accessibility and clarity | ✓, ~ | ✓ | ✓ | ✓ | ✓ | ✓ | ✓ | ✓ |

Table 2. Summary of Themes Across ASA EG and ESSC CoP

*Target audiences:* See Table 2. The ESSC European Code of Statistical Practice was developed in partnership with national statistical offices within the membership of the European Union and the European Free Trade Association and Eurostat to foster adherence to high quality production and dissemination of national and regional official statistics.

*Areas of strong alignment:* Strongest alignment occurs within the institutional environment guidance. There is complete alignment of ASA EG with CoP principle 1 (*professional independence*) and 4 (*commitment to quality*). Both of these responsibilities were aligned with all eight ASA EG principles (pattern denoted by green shading). There is also strong alignment of ASA EG with CoP principles 6 (*impartiality and objectivity*), 7 (*sound methodology*) and 8 (*appropriate statistical procedures*). CoP 15 (*accessibility and clarity*) also has strong alignment across the ASA EGs, although consideration for the validity and interpretability of outputs (ASA EG A.2) should be kept in mind when fulfilling CoP 15. (See Appendix, Detail Table 2.)

*Areas of potential tension:* There are also areas of potential tension, particularly in the area of statistical output guidance. Implementation of CoP 11 (*relevance*), 13 (*timeliness*), and 14 (*comparability*) could conflict with several ASA EG, particularly A (*accountability*), B (*integrity*), C (*stakeholders*), D (*subjects*), and G (*leadership*). As was found in Park and Tractenberg (2023), tensions arising from tight production schedules and stakeholders interests can pressure national statistical offices in their production of accurate and objective statistics; this phenomenon is well-known to professional statisticians, and the tensions observed are not unique to the CoP. Rather, by identifying possible sources of tension, best practices can be leveraged to support informed choices. As a comprehensive set of practices, the ASA EG may be particularly useful to the ethical practitioner seeking balanced implementation.

*Guidance gap:* There are some notable gaps across guidances, particularly with regard to the area of statistical processes. The ASA EG do not provide much guidance regarding CoP 3 (*resources*), CoP 9 (*burden*), or CoP 10 (*cost*). Because the ASA EG are intended for all practitioners, and do not focus on government practitioners where these considerations might be more prevalent, public burden and cost are less salient in the ASA EGs. We also note that CoP does not provide much guidance regarding ASA EG F (*other practitioners*) or H (*misconduct*). The CoP focus includes independent and professional practice, but also emphasizes matters that are most salient (and perhaps less sensitive) to national statistical offices. This CoP focus could potentially lead to a guidance gap for a) those who do not identify as "professional statisticians", for whom "professional independence" may not seem applicable; and b) those who work in statistics and data science at the national level, but do so *outside* of national statistics offices. The ASA EG do not have this guidance gap.

*Discussion:* Overall, there is mixed alignment of the ASA EG with CoP. Alignment is strongest in the area of institutional environment - which includes a professional context that prioritizes independence, somewhat silent in the area of statistical practices, and has the majority of its potential tension in the area of statistical output. Such tensions are familiar to professional statisticians, and are not unique to CoP. The ASA EG may be useful in resolving these potential tensions, particularly for statistical practitioners outside of "official statistics offices" and for those who work on teams with statistical practitioners from diverse groups, organizations, or disciplines. Because of the diversity of ASA membership, apparent gaps between the ASA EG and CoP are not surprising. Alignment could be strengthened through clarification of CoP statistical output guidance that ensures practitioners who follow CoP guidance do not inadvertently violate ASA EGs.

### 3. Alignment of ASA Ethical Guidelines and OECD Good Statistical Practice

*Target audiences:* See Table 3. Recall that the OECD GSP was developed as a 'blueprint' for a strong national statistical system; the blueprint is undergirded by a series of practices intended to provide a framework for national implementation. Although the OECD is a private organization, the GSP is promoted as an assessment tool to evaluate the strength of official statistical systems among member and ascending countries. Thus, the OECD GSP may affect country interests in ways that other international guidance, such as the ISI DPE, does not. There may be greater pressure to demonstrate adherence to the OECD GSP, particularly for countries seeking to ascend within OECD membership, as a result. The set of "good statistical practices" can seem to be a checklist (and operates as such as an assessment tool), which is not the purpose or function of the ASA EGs.

*Areas of strong alignment:* There is strong alignment of the GSP with the ASA EGs. In particular, there is complete alignment of the ASA EG with GSP 2 (*independence*), 6 (*impartiality*), 7 (*methodology*), and 8 (*quality*). There is also notable alignment with GSP 1 (l*egal framework*), and 9 (*access*), 10 (*coordination*) and 11 (*cooperation*). (See Appendix Detail Table 3.)

*Areas of potential tension:* There are few areas of potential tension noted in GSP 12 (*new sources and methods*). Care should be taken to ensure that pursuit of new data sources and methods is not accomplished without respecting ASA EG A (*accountability*), and D (*subjects*) interests.

| OECD Good Statistical Practice (ASA Ethical Guidelines) | A Accountability | B Integrity | C Stakeholder | D Subjects | E Other Disciplines | F Other Statisticians | G Leadership | H Misconduct |
|---|---|---|---|---|---|---|---|---|
| 1 A clear legal and institutional framework | ✓ | ✓ | ✓ | ✓ | ✓ | | ✓ | ✓ |
| 2 Professional independence | ✓ | ✓ | ✓ | ✓ | ✓ | ✓ | ✓ | ✓ |
| 3 Adequacy of resources | (✓) | | | | | | | |
| 4 Protection of privacy | | | ✓ | ✓ | | | | |
| 5. Right to access administrative sources | ✓ | | | ✓ | ✓ | | | ✓ |
| 6. Impartiality, objectivity, and transparency | ✓ | ✓ | ✓ | ✓ | ✓ | ✓ | ✓ | ✓ |
| 7. Sound methodology and professional standards | ✓ | ✓ | ✓ | ✓ | ✓ | ✓ | ✓ | ✓ |
| 8. Quality of statistical outputs and processes | ✓ | ✓ | ✓ | ✓ | ✓ | ✓ | ✓ | ✓ |
| 9. User-friendly access and commitment to respond to misinterpretation by users | ✓ | ✓ | ✓ | ✓ | ✓ | ✓ | | |
| 10. Coordination of statistical activities | ✓ | ✓ | ✓ | ✓ | ✓ | ✓ | ✓ | |
| 11. International cooperation | ✓ | ✓ | ✓ | ✓ | ✓ | ✓ | ✓ | |
| 12. Explore new, alternative data sources/methods | ✓,~ | ✓ | ✓ | ✓,~ | | | | ✓ |

Table 3. Summary of Themes Across ASA EG and OECD GSP

*Guidance gap:* The ASA EG has limited alignment with the GSP 3 (*resources*); this is not unexpected given the GSP's focus on national statistical offices, compared to the more practitioner-specific, context-agnostic, guidance from the ASA EGs. Although only two of the ASA EG Principles relate to GSP 4 (*privacy*), the guidance from ASA EG

Principle D is detailed. The considerations for Research Subjects/Data Subjects and Those Affected by Statistical Practices (Principle D) and Responsibilities to Stakeholders (Principle C) go far beyond privacy, which is a gap in guidance that is shared across all of the official statistics guidance documents.

*Discussion:* Overall, there is strong alignment of the ASA EG and the GSP, with few areas of potential tension in implementation. These arise in GSP 12 (exploring new, alternative data sources and methods), principally because the ASA EG Principles of Accountability (A) and Responsibilities to Research Subjects, Data Subjects, or Those Directly Affected by Statistical Practices (D) require careful consideration of any use to which data are put (even if alternative methods are under consideration).

### 4. Alignment of ASA Ethical Guidelines and ISI Statement of Ethics

*Target audiences:* See Table 4. Recall that, like the ASA, the ISI is a non-profit, non-governmental organization whose mission is to support professional statisticians internationally. Rather than an ethical practice standard, the ISI DPEtakes the form of strategies for the ethical practitioner; it acts to communicate the ISI's support of professional statisticians, note concern to national authorities and international agencies, and to collaborate on joint communications with other professional societies to amplify messages.

*Areas of strong alignment:* There is strong alignment of the ASA EG with the ISI DPE. As denoted by green shading, there is complete alignment of the ASA EG with DPEValue 3 (t*ruthfulness, independence, transparency*), DPEPrinciples 1 (*objectivity, transparency*), 4 (*avoid conflict of interests*), 5 (*avoid preempted outcomes*), 8 (*confidence*), 9 (*review methods and findings*), and 11 (*responsibility*). We also observe strong alignment of ASA EG Principle A (*accountability*) with the entirety of the DPE(denoted by green shading). It is also interesting to note that every element of ASA EG E (*responsibilities to other disciplines*) is aligned with both DPEPrinciples 1 (*objectivity*) and 8 (*confidence*), denoted by dark green shading. (See Appendix, Detail Table 4.) The strength of alignment is expected given that both organizations seek to promote independent, valid, and reproducible statistical outputs.

International Guidelines align with ASA Ethical Guidelines

| ISI Statement of Ethics / ASA Ethical Guidelines | A Accountability | B Integrity | C Stakeholder | D Subjects | E Other Disciplines | F Other Statisticians | G Leadership | H Misconduct |
|---|---|---|---|---|---|---|---|---|
| **Shared Professional Values** | | | | | | | | |
| 1 Respect for data providers, communities, and others' work | ✓ | (✓) | ✓ | ✓ | | ✓ | ✓ | |
| 2 Professionalism: responsibility; competence; and informed judgement | ✓ ~ | ✓ | ✓ | | ✓ | ✓ | ✓ | ✓ |
| 3 Truthfulness and integrity: independence; objectivity; and transparency | ✓ | ✓ | ✓ | (✓) | ✓ | (✓) | ✓ | |
| **Ethical Principles** | | | | | | | | |
| 1. Pursue objectivity by transparent methods and communication | ✓ | ✓ | ✓ | ✓ | ✓ | ✓ | ✓ | ✓ |
| 2. Clarify obligations and roles by competency areas | ✓ | | ✓ | ✓ | ✓ | ✓ | ✓ | ✓ |
| 3. Assess alternatives impartially | (✓) | ✓ | ✓ | (✓) | (✓) | ✓ | (✓) | |
| 4. Avoid conflicting interests | ✓ | ✓ | ✓ | (✓) | ✓ | ✓ | ✓ | (✓) |
| 5. Avoid preempted outcomes | ✓ | ✓ | ✓ | (✓) | ✓ | ✓ | ✓ | ✓ |
| 6. Guard privileged information | (✓) | (✓) | (✓) | ✓ | (✓) | | | (✓) |
| 7. Exhibit professional competence | ✓ | | | (✓) | | (✓) | ✓ | |
| 8. Maintain confidence in statistics | (✓) | ✓ | ✓ | ✓ | ✓ | (✓) | ✓ | (✓) |
| 9. Expose and review methods and findings | ✓ | ✓ | ✓ | ✓ (~) | (✓) | ✓ | ✓ | (✓) |
| 10. Communicate ethical principles | ✓ | | (✓) | (✓) | ✓ | ✓ | ✓ | ✓ |
| 11. Be responsible for the integrity of the discipline | ✓ | ✓ | ✓ | ✓ | ✓ | ✓ | ✓ | ✓ |
| 12. Protect the interests of subjects | (✓) | (✓) | ✓ | ✓ | (✓) | | ✓ | |

Table 4. Summary of Themes Across ASA EG and ISI SOE

*Areas of potential tension:* There are very few areas of potential tension observed. We note that when *describing methods and findings* (DPE9), care should be taken to protect *subjects' interests* regarding privacy at the individual and group level (ASA EG D). There are two elements of DPEValue 2 (*professionalism*) where there is tension. In the first instance, DPEValue 2 begins, "We use our statistical knowledge, data, and analyses for the Common Good to serve the society." This is an expression of a utilitarian perspective (where the common good is prioritized); the ASA EGs take a virtue perspective, where "the ethical statistical practitioner" is the target- without specifying that this particular practitioner has "the common good" as an objective. In the second instance, DPEValue 2 continues, "We…work to change laws we believe impede good statistical practice". ASA EG element A11 states that the ethical statistical practitioner "Follows applicable policies, regulations, and laws relating to their professional work, unless there is a compelling ethical justification to do otherwise." Although both guidelines agree to the idea that practitioners should work to change laws where they impede good statistical practice, individual practitioners may have differing interpretations of compelling justification to do so. Also, ASA EG Principle G and the Appendix (not shown in these summary tables, but included in the detailed tables) include specific responsibilities for leaders (G) and organizations that employ statistics practitioners or their outputs (Appendix) to promote policies that uphold practitioner independence and to ensure an ethical workplace. In this way, the ASA EGs may be seen as diffusing the responsibility to "work to change laws" to include policies and workplace procedures, so as to promote "good statistical practice" (ISI) and "ethical statistical practice" (ASA).

*Guidance gap:* We do not observe significant guidance gaps, apart from the extent of guidance for leaders (ASA EG G) and organizations (ASA EG Appendix) that goes beyond what the individual practitioner can and should do to promote both"good statistical practice" (ISI) and "ethical statistical practice" (ASA). However, there is one shared professional value (SPV 2) which includes something that might be relevant in an international context but is explicitly not applicable in a US context. The ISI SPV of Professionalism asserts that statistical practitioners "are responsible for the fitness of data." This aspect is not included in the ASA EG, and could create a legal liability for practitioners in the US. Specifically, ASA EG outline that the ethical obligation is to *use only data that are fit for purpose* - i.e., to establish that the data are fit for the intended purpose, otherwise, it cannot be used. The ISI element charges the responsibility for the data's fitness for purpose to the statistician; but in the ASA EG context, where possibly most statistical practitioners do not collect the data, they cannot be, nor be held, responsible for its fitness for purpose.

*Discussion:* Two documents from two professional statistical organizations should align closely, and these do. There is strong alignment between the ASA EG and the ISI DPE. Because the DPEare intended as roles/strategies for ethical conduct of statisticians in particular (generally, communication) situations, the ISI DPEcould be seen as a helpful complement to the practice-specific guidance in the ASA EG. Tension does arise between these guidances primarily from a focus on the common good and society (ISI), which are priorities for national statisticians (i.e., a core constituency of the ISI) but not for academic and industry/business applications (i.e., core constituencies of the ASA). The ASA EGs are also possibly more relevant for statistical practitioners in government settings who do not specifically work in "national statistics" like census and other designated statistical agencies.

## 5. *Alignment among international guidances*

The preceding section described the alignment of several international guidelines with the ASA EGs. Overall, we found strong alignment with most of these guidelines, suggesting the value of the (detailed) ASA EG as a practical resource to promote ethical conduct and avoid potential tensions. Specifically, the preceding tables show that every statistical practitioner, whether in a national/federal statistics office or environment or in other international statistical practice context, can follow the ASA EGs and be confident that they will also not violate the specific guidance of the UN, CoP, OECD, or ISI.

This analysis also makes apparent the wide range of guidelines for the professional statistician and data scientist working within the specific official statistics contexts internationally to consider. We therefore next examined the alignment among these international guidances to identify commonalities and areas of difference.

### a. *Alignment of UN Fundamental Principles of Official Statistics and ESSC European Statistical Code of Practice*

*Target audiences*: See Table 5a. Recall that the UN FPOS and the ESSC CoP are intended to describe core features of national statistical systems and the production of official statistics. By their nature, the regional audience (CoP) is nested in the larger (UN) international audience. Given that the European Union nation states that contribute their data to UN resources and reports coordinate the production and dissemination of statistics with Eurostat, one can anticipate guidance from EU and UN national level statistics groups to align closely.

*Areas of strong alignment*: Alignment is strongest in the area of the institutional environment. In particular, we note strong alignment of the FPOS with CoP 4 (*quality*), 5 (*confidentiality*), and 6 (*objectivity*). (See Appendix, Detail Table 5a.)

*Areas of potential tension*: We also note areas of potential tension between CoP and FPOS, particularly with CoP 1bis (*coordination*) and 2 (*mandate for data collection and access*). Similarly, we note potential tension of CoP with FPOS 10 (*cooperation*). These are shown as a red tilde within parentheses (~) to denote indirect sources of potential tension, which could be addressed with clarifying language.

| ESSC European Code of Statistical Practice | 1 Make data publicly available | 2 Collect using scientific methods | 3 Present using professional standards | 4 Comment on misuse | 5 Choose data considering quality, timeliness, cost, and respondent burden | 6 Use confidential data only for statistical purposes | 7 Operate the statistical system within a legal framework | 8 Coordinate across national statistical agencies | 9 Use international statistical classifications consistently | 10 Cooperate across national statistical systems |
|---|---|---|---|---|---|---|---|---|---|---|
| **Institutional Environment** | | | | | | | | | | |
| 1 Professional independence | | ✓ | | | | ✓ | | ✓ | | ~ |
| 1bis Coordination and cooperation | | (~) | | (~) | | (~) | (~) | (~) | | ✓ |
| 2 Mandate for data collection and access to data | (✓) | | ✓ | ✓ | ✓ | | (✓) | | | ✓ |
| 3 Adequacy of resources | | | ✓ | ✓ | ✓ | | | | | ✓ |
| 4 Commitment to quality | | ✓ | | ✓ | | ✓ | ✓ | ✓ | ✓ | |
| 5. Statistical confidentiality and data protection | ✓ | ✓ | (✓) | ✓ | ✓ | | ✓ | ✓ | | (~) |
| 6. Impartiality and objectivity | | ✓ | ✓ | | ✓ | ✓ | ✓ | ✓ | | (~) |
| **Statistical Processes** | | | | | | | | | | |
| 7 Sound methodology | | ✓ | ✓ | | ✓ | | | | | |
| 8 Appropriate statistical procedures | | ✓ | ✓ | ✓ | | | | | | |
| 9 No undue respondent burden | | | | | ✓ | | | | | |
| 10 Cost effectiveness | | | | | ✓ | | | | | |
| **Statistical Output** | | | | | | | | | | |
| 11 Relevance | | | ✓ | | ✓ | | | | | |
| 12 Accuracy and reliability | | ✓ | ✓ | (✓) | | | | | | |
| 13 Timeliness and punctuality | | | | | ✓ | | | | | |
| 14 Coherence and comparability | | | | | ~ | | | ✓ | ✓ | ✓ |
| 15 Accessibility and clarity | ✓ | ✓ | ✓ | ✓ | | | | ✓ | ~ | ✓ |

Table 5a. Summary of Themes Across UN FPOS and ESSC CoP

We were somewhat surprised to observe the potential for tension arising from CoP 1bis (*coordination and cooperation*) for FPOS 2 (*scientific data collection*); 4 (*comment on misuse*); 6 (*solely statistical purposes of data*); 7 (*public laws/regulations*); and 8 (*coordination within countries*). This arises from the specificity of the ESSC institutional environmental characteristics and the more general descriptions of practices in FPOS, and could be addressed with clarifying language.

*Guidance gap:* No guidance gaps were observed, though the commitment to FPOS 9 (*consistent use of international classifications*) could be amplified in the CoP. Perhaps this is implicit in daily practice among EU countries. The emphasis in the FPOS on collecting data using scientific methods (2) and presenting using professional standards (3), and that in the CoP on sound methodology (7) and appropriate statistical procedures (8) ensure that any guidance gaps will not result in lower quality outputs.

*Discussion:* We observed less alignment than we anticipated among these international guidances, despite the similarities in their target audiences. There is no obvious way to resolve the tensions noted by consulting the two guidances alone. Instead, we suggest ASA EGs, given their detailed and wide scope, can be used to resolve these tensions, and/or for suggestions that clarify language that leads to tensions.

b. *Alignment of UN Fundamental Principles of Official Statistics and OECD Good Statistical Practice*

| UN Fundamental Principles of Official Statistics<br><br>OECD Good Statistical Practice | 1 Make data publicly available | 2 Collect using scientific methods | 3 Present using professional standards | 4 Comment on misuse | 5 Choose data considering quality, timeliness, cost, and respondent burden | 6 Use confidential data only for statistical purposes | 7 Operate the statistical system within a legal framework | 8 Coordinate across national statistical agencies | 9 Use international statistical classifications consistently | 10 Cooperate across national statistical systems |
|---|---|---|---|---|---|---|---|---|---|---|
| 1 A clear legal and institutional framework | | | | | | | ✓ | | | |
| 2 Professional independence | | (✓) | | | ✓ | | | ~ | ~ | ~ |
| 3 Adequacy of resources | | | | | | | | (✓) | | |
| 4 Protection of privacy | | | | | | ✓ | | | | |
| 5. Right to access administrative sources | ✓ | | | | ✓ | ✓ | | (✓) | (✓) | (✓) |
| 6. Impartiality, objectivity, and transparency | ✓ | ✓ | ✓ | ✓ | | | | | | |
| 7. Sound methodology and professional standards | ✓ | ✓ | ✓ | ✓ | | | | | | |
| 8. Quality of statistical outputs and processes | | ✓ | ✓ | ✓ | | | | | | |
| 9. User-friendly access and commitment to respond to misinterpretation by users | ✓ | | | ✓ | | | | | | |
| 10. Coordination of statistical activities | | | | | | | | ✓ | ✓ | ✓ |
| 11. International cooperation | | | | | | | | | ✓ | ✓ |
| 12. Explore new, alternative data sources/methods | ✓ | ✓ | ✓ | ✓ | | | | | | |

Table 5b. Summary of Themes Across UN FPOS and OECD GSP

*Target audiences: See Table 5b.* The OECD GSP were written specifically with the UN FPOS in mind. Similar to section 5a (above), the UN FPOS and the OECD GSP have nested audiences. However, since membership and ascension into the OECD is ostensibly open to any nation state, the 'nesting' occurs across geographic areas. Nonetheless, nation states of the "global north" (that is, countries with more developed economies) predominate OECD membership. Recall also that OECD GSP are promoted as an assessment tool, and therefore may be used, and can affect countries' interests, differently than the UN FPOS. Finally, in the documentation, the OECD GSP are described as having come into being in response to UN FPOS. Therefore, unlike the expected alignment of UN and EU guidance, the UN and OECD guidances might be expected to diverge or cover different aspects of statistical practice.

*Areas of strong alignment:* Not surprisingly, we observe alignment of FPOS 1 (*public data*), 2 (*scientific collection*), 3 (*professional standards*), and 4 (*comment on misuse*) with GSP 6 (*impartiality*), 7 (*sound methods*), 8 (*quality outputs*), and 12 (*explore data sources/methods*). We also note the alignment of FPOS 8 (*coordination),* 9 (*international classification*), and 10 (*cooperation*) with GSP 10 (*coordination*) and 11 (*cooperation*).

*Areas of potential tension:* FPOS 8 *(coordination)*, 9 *(international classification),* and 10 *(cooperation)* can create tension with GSP 2 *(professional independence).* This was also noted in our examination of alignment of FPOS with the ASA EGs.

*Guidance gap:* In general, there are no gaps in guidance, *per se.*

*Discussion:* Of the guidelines examined in this paper, only FPOS does not feature professional independence. he ASA EGs are specifically intended, as noted earlier, for any and all statistical practitioners, which as we have noted elsewhere, makes them useful and important for promoting ethical statistical practice across all contexts. However, the idea of "professional independence" might not mean the same thing to people with the job title/training "statistician" or "data scientist" as they do for people with different backgrounds and job titles who nevertheless contribute to statistical outputs across official statistics contexts. Guidelines providing detailed practices can be especially helpful to practitioners, particularly those who use statistical practices but are outside of statistical agencies, or who do not have the title or training to consider "professional independence" to be applicable to themselves.

   c. *Alignment of UN Fundamental Principles of Official Statistics and ISI Statement of Ethics*

*Target audiences:* See Table 5c. Recall that the target audiences of FPOS and DPEare both broad geographically. However, FPOS directs its guidance to national statistical offices, and indirectly to their employees in the production of official statistics. While the ISI is not specifically limited to national statistical offices or practitioners, DPE's guidance is intended for individuals, and to facilitate communication across national statistical offices in the adherence of commonly-held norms.

*Areas of strong alignment:* Importantly, while there is abundant evidence of alignment between DPEand FPOS, much of the alignment is indirect - that is, the indicators are in parenthesis wherever there is no direct match in the language of FPOS and DPEarticulation. This might be due to the fact that the DPEis intentionally aspirational and also targeted to the practitioner, while FPOS was intended to characterize the national statistical system and might be less aspirational, although implementation guidance has not yet been completed. The least ambiguous alignment is between FPOS 2 (*scientific data collection*) and the majority of the DPEelements.

*Areas of potential tension:* Similar to our observations in Table 5b examining the alignment of the FPOS with the ASA EGs, the UN principles of *coordination* (FPOS 8) and *cooperation* (spanning both FPOS 9 and 10) have the potential to create conflicts with most of the ISI DPEelements. The independence of the practitioner is a core element throughout the ISI DPE, while this is absent from the FPOS. Conversely, the FPOS indirectly support independence (such as, through reliance on professional standards, FPOS 3)), but place explicit priority on cooperation and coordination.

| UN Fundamental Principles of Official Statistics / ISI Statement of Ethics | 1 Make data publicly available | 2 Collect using scientific methods | 3 Present using professional standards | 4 Comment on misuse | 5 Choose data considering quality, timeliness, cost, and respondent burden | 6 Use confidential data only for statistical purposes | 7 Operate the statistical system within a legal framework | 8 Coordinate across national statistical agencies | 9 Use international statistical classifications consistently | 10 Cooperate across national statistical systems |
|---|---|---|---|---|---|---|---|---|---|---|
| **Shared Professional Values** | | | | | | | | | | |
| 1 Respect for data providers, communities, and others' work | | ✓ | (✓) | ✓ | | (✓) | | (~) | | (~) |
| 2 Professionalism: responsibility; competence; and informed judgement | (✓) | ✓ | | ✓ | ✓ | (✓) | | (~) | (✓) | (~) |
| 3 Truthfulness and integrity: independence; objectivity; and transparency | | ✓ | ✓ | ✓ | | | | (~) | (✓) | (~) |
| **Ethical Principles** | | | | | | | | | | |
| 1. Pursue objectivity by transparent methods and communication | | ✓ | ✓ | | | | | (~) | ✓ | (~) |
| 2. Clarify obligations and roles by competency areas | | ✓ | | | | | | (✓) | | (✓) |
| 3. Assess alternatives impartially | | ✓ | (✓) | | | | | | (✓) | |
| 4. Avoid conflicting interests | | | | | (✓) | (✓) | | (~) | | (~) |
| 5. Avoid preempted outcomes | | ✓ | | (✓) | | | | (~) | | |
| 6. Guard privileged information | | | | | | (✓) | | (~) | (✓) | (~) |
| 7. Exhibit professional competence | | ✓ | | | | | | | (✓) | |
| 8. Maintain confidence in statistics | (✓) | (✓) | ✓ | (✓) | | | | (~) | ✓ | (~) |
| 9. Expose and review methods and findings | | (✓) | (✓) | (✓) | | | | (✓) | | ✓ |
| 10. Communicate ethical principles | | | | (✓) | | | | (✓) | ✓ | ✓ |
| 11. Be responsible for the integrity of the discipline | | ✓ | (✓) | ✓ | | | | | (✓) | (~) |
| 12. Protect the interests of subjects | (✓) | | | | | (✓) | | (~) | | (~) |

Table 5c. Summary of Themes Across UN FPOS and ISI SOE

*Guidance gap:* There is a clear gap between FPOS 7 (legal framework) and the DPE, where there is no overlap. This is anticipated given the target audiences of the documents.

*Discussion:* As observed when examining alignment across the ASA EG and UN FPOS, there are consistent tensions between guidances that explicitly recognize *independence* (such as ASA EG and ISI DPE) and those that emphasize *collaboration* and *coordination* (such as UN FPOS). It is important to consider the different guidance offered for organizations and individuals, and also to anticipate where organizational objectives and individual ethical practice may inadvertently conflict.

### d. Alignment of OECD Good Statistical Practice and ESSC European Statistical Code of Practice

*Target audiences:* See Table 5d. The GSP and CoP come from organizations with considerable geographic overlap. Both are used in assessing the quality of national statistical offices and their products for policy making purposes. There may be program implications of these assessments. For example, Eurostat could cite adherence to CoP when evaluating EU member official statistics. In the case of OECD, iIt is possible that assessments using the GSP could be used in consideration of development or other program funding.

| ESSC European Code of Statistical Practice / OECD Good Statistical Practice | 1 A clear legal and institutional framework | 2 Professional independence | 3 Adequacy of resources | 4 Protection of privacy | 5 The right to access administrative sources | 6 Impartiality, objectivity, and transparency | 7 Sound methods & professional standards | 8 Quality of statistical outputs & processes | 9 User-friendly access and commitment to respond to user misinterpretation | 10 Coordination of statistical activities | 11 International cooperation | 12 Explore new and alternative data sources and methods |
|---|---|---|---|---|---|---|---|---|---|---|---|---|
| **Institutional Environment** | | | | | | | | | | | | |
| 1 Professional independence | | ✓ | | | | ✓ | | ✓ | | ~ | ~ | |
| 1bis Coordination and cooperation | | (~) | | (~) | | (~) | (~) | (~) | | ✓ | ✓ | |
| 2 Mandate for data collection and access to data | (✓) | | ✓ | ✓ | ✓ | | (✓) | | | ✓ | ✓ | |
| 3 Adequacy of resources | | | ✓ | ✓ | ✓ | | | | | ✓ | ✓ | |
| 4 Commitment to quality | | ✓ | | ✓ | | ✓ | ✓ | ✓ | ✓ | | | ✓ |
| 5 Statistical confidentiality and data protection | ✓ | ✓ | (✓) | ✓ | ✓ | | ✓ | ✓ | | (~) | (~) | |
| 6 Impartiality and objectivity | | ✓ | | | ✓ | ✓ | ✓ | ✓ | | (~) | (~) | (✓) |
| **Statistical Processes** | | | | | | | | | | | | |
| 7 Sound methodology | | ✓ | | (✓) | | ✓ | ✓ | ✓ | ✓ | | (✓)(~) | (~) |
| 8 Appropriate statistical procedures | | ✓ | | (✓) | | ✓ | ✓ | ✓ | ✓ | (~) | (✓)(~) | (~) |
| 9 No undue respondent burden | (✓) | | | | | | (✓) | | (✓) | | | ✓ |
| 10 Cost effectiveness | (✓) | (~) | (✓) | | | | | (✓) | | ✓ | ✓ | |
| **Statistical Output** | | | | | | | | | | | | |
| 11 Relevance | (✓) | (~) | | (~) | | | ✓ | (✓) | | (✓)(~) | (✓)(~) | (✓)(~) |
| 12 Accuracy and reliability | | ✓ | | | | ✓ | ✓ | ✓ | | (✓)(~) | (✓)(~) | (✓) |
| 13 Timeliness and punctuality | | (~) | (~) | (~) | | | | (✓) | | (✓)(~) | (✓)(~) | |
| 14 Coherence and comparability | | (✓)(~) | | | (✓) | (✓)(~) | ✓ | ✓ | ✓ | | (✓) | (✓)(~) |
| 15 Accessibility and clarity | (✓) | | | | | ✓ | ✓ | ✓ | ✓ | (✓)(~) | (✓)(~) | |

Table 5d. Summary of Themes Across ESSC CoP and OECD GSP

*Areas of strong alignment:* As expected, we observed alignment across these guidances. This was most apparent in support of OECD GSP 7 (*methods and professional standards*) and 8 (*quality outputs*) with exceptions noted below. However, although each of the principles in the GSP and CoP is supported by at least one element (often several) in the other guideline, it is notable that the overlap is not as cohesive as we anticipated given target audiences. Further, this overlap seems to be a mix of direct and indirect (indicated by checkmarks with and without parentheses, respectively). The UN FPOS do not highlight professional independence, but OECD GSP and ESSC CoP both do.

*Areas of potential tension:* We also observe many areas of potential tension, particularly in the area of statistical outputs. Both the GSP (10) and (11) and CoP (1bis) and (14) identify principles of *collaboration* and *cooperation*. These principles can be particularly challenging to implement in practice. *Cooperation* (GSP 11) can lead to or create tension with commitments to *relevance* (CoP 11) and *timeliness* (CoP 13), for example. *Collaboration* (GSP 10) can affect *accessibility and clarity* (CoP 15). Of course, collaboration and coordination can also put positive pressures on adherence to other shared principles; therefore, the tensions themselves should not be interpreted as conflict *per se*, but as opportunities where best practices can be clarified. Also important, we observed many instances where both alignment (black) and potential tension (red) arise in the same cell. This could be particularly problematic in two ways. Firstly, it is possible that each guidance document has some internal discord that could be simply confusing but could also be self-contradictory. Secondly, if one group tries to follow both guidances, or if team members seek to each follow one of these guidances, there could be tensions that are not easily resolved.

*Guidance gap:* We do not observe any guidance gaps. Rather than the gaps, the more important observation is the yellow highlighted cells with both alignment and tension identified.

*Discussion:* We were surprised to observe the number of tensions identified. Paradoxically, although both guidances emphasized principles of cooperation and collaboration (and are therefore aligned), these principles can be among the most challenging to implement in balance with other (shared) principles, such as relevance and timeliness. As shown in Table 5d, this occurs within each the CoP and GSP, and, additionally, when attempting to adhere jointly to both guidances. Nonetheless, these tensions can also provide positive pressure to adopt best practices.

### e.   Alignment of OECD Good Statistical Practice and ISI Statement of Ethics

*Target audiences:* See Table 5e. Recall that OECD GSP and ISI DPEare both guidelines developed by international nongovernmental organizations to support quality in statistical practice. The guidelines do not carry the weight of national statistical laws or regulations; instead, the guidelines are intended to support norms and communication. However, OECD GSP have been promoted to facilitate assessment of national statistical offices, and therefore could have substantial impact on national statistical policy. The ISI DPE, intended for individuals, have lower likelihood of impacting national statistical policy (since this is not their explicit intention).

*Areas of strong alignment:* There are patterns of direct and indirect alignment of OECD GSP and ISI DPE. There is alignment of ISI DPEwith OECD GSP 2 (*independence*), 6 (*objectivity*), and 12 (*new methods*) but this is substantially (if not generally) indirect. We also observe patterns of alignment of the OECD GSP with ISI DPEEP 1 (*objectivity*) but also note areas of potential tension (see further below).

*Areas of potential tension:* We observe notable patterns of potential tension. Specifically, implementing OECD 11 (*coordination*) and 12 (*cooperation*) could cause tension when implementing nearly all of the ISI DPE. This is a consistent challenge where cooperation and collaboration are specified in guidelines, because while collaboration and cooperation can apply positive pressure for coherence in statistical practice in ways that can foster critical thinking and best practices, requirements for cooperation and collaboration can also create pressures to cut corners or fail to follow the ethical practice standards for statistics. Without specific guidance about how to balance these practices, tension remains.

*Guidance gap:* We observe a lack of guidance alignment particular to OECD GSP 1 (*legal framework*), and 5 (*data access rights*). This may relate to the nature of ISI DPEas being a non-governmental organization intended to meet the needs of individual statistical practitioners (foremost), rather than primarily national statistical offices. The guidance need not be contradictory; given that independence and impartiality are prioritized in the OECD GSP (#2, 6, 7, 8) and DPE(SPV2, SPV3; EP 1, 3, 5, 7, 8, and 11), language promoting a clear legal framework for national statistical offices is generally compatible with communication guidance for national statistical offices, particularly in times where impartiality may be challenged. Since the targets of these guidances are diverse, this sort of gap in guidance is expected, but the alignment towards integrity can offset the gap.

 *Discussion:* Overall, we observe less alignment than anticipated across OECD GSP and ISI DPE. In part, this may be due to differences in the intended audiences and uses of guidance documents. Nonetheless, the emphasis on *coordination* (10) and *cooperation* (11) within the OECD GSP presents tensions in achieving other OECD GSP and most of the ISI DPE. Where specific best practices can be provided, and whenever independence and ethical practice are prioritized, professional statisticians will be in a stronger position to balance potentially competing guidance.

| ISI Statement of Ethics / OECD Good Statistical Practice | 1 A clear legal and institutional framework | 2 Professional independence | 3 Adequacy of resources | 4 Protection of privacy | 5 The right to access administrative sources | 6 Impartiality, objectivity, and transparency | 7 Sound methods & professional standards | 8 Quality of statistical outputs & processes | 9 User-friendly access and commitment to respond to user misinterpretation | 10 Coordination of statistical activities | 11 International cooperation | 12 Explore new and alternative data sources and methods |
|---|---|---|---|---|---|---|---|---|---|---|---|---|
| **Shared Professional Values** | | | | | | | | | | | | |
| 1 Respect for data providers, communities, and others' work | | | | ✓ | | (✓) | | | | (~) | (~) | |
| 2 Professionalism: responsibility; competence; and informed judgement | | ✓ | (✓) | | | ✓ | ✓ | | (✓) | (~) | (~) | (✓) |
| 3 Truthfulness and integrity: independence; objectivity; and transparency | | ✓ | | | | ✓ | (✓) | | | (~) | (~) | (✓) |
| **Ethical Principles** | | | | | | | | | | | | |
| 1. Pursue objectivity by transparent methods and communication | | ✓ | (✓) | | | ✓ | ✓ | ✓ | ✓ | (~) | (~) | (✓) |
| 2. Clarify obligations and roles by competency areas | | (✓) | | | | | | | | (✓) | (~) | |
| 3. Assess alternatives impartially | | (✓) | (✓) | | | ✓ | | | | | | ✓ |
| 4. Avoid conflicting interests | | (✓) | | | | | | | (✓) | (~) | (~) | |
| 5. Avoid preempted outcomes | | | | | | (✓) | | | | (~) | (~) | (✓) |
| 6. Guard privileged information | | | | ✓ | | | | | | (~) | (~) | |
| 7. Exhibit professional competence | | | | | | | (✓) | | | | | ✓ |
| 8. Maintain confidence in statistics | | (✓) | (✓) | | | (✓) | | ✓ | (✓) | (~) | (~) | ✓ |
| 9. Expose and review methods and findings | | (✓) | | | | | | | (✓) | (✓) | ✓ | ✓ |
| 10. Communicate ethical principles | | (✓) | | | | | | | | (✓) | ✓ | |
| 11. Be responsible for the integrity of the discipline | | (✓) | | | | ✓ | | | ✓ | | (~) | (✓) |
| 12. Protect the interests of subjects | | | | (✓) | | | | | | (~) | (~) | |

Table 5e. Summary of Themes Across OECD GSP and ISI SOE

*f. Alignment of ESSC European Statistical Code of Practice and ISI Statement of Ethics*

*Target audiences:* See Table 5f. Recall that the ESSC CoP and the ISI DPEwere developed for very different purposes and audiences. The CoP were developed to guide national statistical offices in preparing official statistics, particularly to coordinate official statistical production for European Union countries. The ISI DPEwere developed as a way to support communication of normative ethical practice across the statistician profession, internationally. Both have informative elements for the individual; the DPEis specifically for the individual while the CoP also offers environmental and contextual guidance (i.e., not at the individual level).

*Areas of strong alignment:* Perhaps reflecting the diverse origins and target audiences for these guidances, we observed less alignment than expected given the international applicability of both. We observe strong alignment of ESSC CoP 4 (*quality*) 6 (*impartiality*) with ISI DPE. We also note direct and indirect alignment of ISI DPEvalue 2 (*professionalism*) and ISI DPEprincipal 1 (*objectivity*), 3 (*impartiality*) with ESSC CoP, but this alignment is mixed with areas of tension (see below).

*Areas of potential tension:* As has been seen whenever professional organization guidance is aligned with national level guidance, areas of tension with ISI DPEare consistently observed between ESSC CoP 1bis (*coordination*), 2 (*data access*) and 13 (*timeliness*). As was observed in Table 5d (ESSC with OECD), we observed many instances where both alignment (black) and potential tension (red) arise in the same cell.

| ESSC European Code of Statistical Practice / ISI Statement of Ethics | 1 Professional Independence | 1bis Coordination and Cooperation | 2 Mandate for Data Collection and Access to Data | 3 Adequacy of Resources | 4 Commitment to Quality | 5 Statistical Confidentiality and Data Protection | 6 Impartiality and Objectivity | 7 Sound Methodology | 8 Appropriate Statistical Procedures | 9 Non-excessive Burden on Respondents | 10 Cost Effectiveness | 11 Relevance | 12 Accuracy and Reliability | 13 Timeliness and Punctuality | 14 Coherence and Comparability | 15 Accessibility and Clarity |
|---|---|---|---|---|---|---|---|---|---|---|---|---|---|---|---|---|
| **Shared Professional Values** | | | | | | | | | | | | | | | | |
| 1 Respect for data providers, communities, and others' work | | (✓) | (✓) | | | ✓ | | | | (✓) | | | | | | |
| 2 Professionalism: responsibility; competence; and informed judgement | (✓) | ✓(✓) | (✓) | (✓) | ✓ | | ✓ | ✓ | ✓ | | | | ✓(✓) | ✓(✓) | | (✓) |
| 3 Truthfulness and integrity: independence; objectivity; and transparency | ✓ | ✓(✓) | (✓)(✓) | | ✓ | | ✓ | | | | | | ✓ | ✓(✓) | | ✓ |
| **Ethical Principles** | | | | | | | | | | | | | | | | |
| 1. Pursue objectivity by transparent methods and communication | ✓ | ✓(✓) | (✓)(✓) | | ✓ | | ✓ | ✓ | ✓ | | | | ✓ | ✓(✓) | | ✓ |
| 2. Clarify obligations and roles by competency areas | | ✓ | ✓ | | | | | | | | | | | | (✓) | |
| 3. Assess alternatives impartially | ✓ | ✓ | | (✓) | ✓ | | ✓ | ✓ | ✓ | (✓) | (✓) | | | | | ✓ |
| 4. Avoid conflicting interests | (✓) | (✓) | | | | | | | | | | | | | | |
| 5. Avoid preempted outcomes | ✓ | ✓(✓) | (✓)(✓) | | ✓ | | ✓ | ✓ | ✓ | | | | ✓(✓) | | | |
| 6. Guard privileged information | | (✓) | | | | ✓ | | | | | | | | | | |
| 7. Exhibit professional competence | | ✓ | | | ✓ | | | | | | | | | | | |
| 8. Maintain confidence in statistics | (✓) | (✓)(✓) | (✓) | | ✓ | | ✓ | | | | | | ✓ | ✓(✓) | | |
| 9. Expose and review methods and findings | (✓) | (✓)(✓) | | | ✓ | | ✓ | | | | | | | ✓(✓) | | ✓ |
| 10. Communicate ethical principles | | ✓ | | | ✓ | (✓) | | | | | | | | | (✓) | |
| 11. Be responsible for the integrity of the discipline | | ✓(✓) | (✓)(✓) | | ✓ | | | | | | | | ✓ | | | |
| 12. Protect the interests of subjects | (✓) | ✓(✓) | | | | ✓ | | | | (✓) | | | (✓) | (✓) | | ✓ |

Table 5f. Summary of Themes Across ESSC ECSP and ISI SOE

*Guidance gap:* We do not note clear gaps in guidance, but frequently alignment is indirect. Also, as was observed in Table 5d, the highlighted cells where both alignment and tension were identified are plentiful.

*Discussion:* As noted previously, where two independent guidances explicitly call for coordination, there is perhaps greater opportunity for tension across guidances. Collaboration can encourage ethical behavior, but can also pressure professionals to seek conformity at the risk of best practice.

### g. Summary Themes Across all Guidelines Examined

| | ASA EG | UN FPOS | ESSC CoP | OECD GSP | ISI SoE |
|---|---|---|---|---|---|
| ASA EG | x | Aligned | Aligned | Aligned | Aligned |
| UN FPOS | | x | Mixed | (Aligned) | Aligned + Tension |
| ESSC CoP | | | x | Mixed | Mixed |
| OECD GSP | | | | x | Aligned + Tension |
| ISI SoE | | | | | x |

Table 5g. Overall Summary of Themes: Alignment Across Examined Guidelines

*Target audiences:* See Table 5g. At a broad level, we describe patterns of alignment and possible tension across the five guidelines examined here. Two of these (ESSC CoP and OECD GSP) pertain to constituencies generally considered the global North. Two (ASA EG and ISI DPE) contain guidance specific for individuals; two (UN FPOS and OECD GSP) are explicitly for organizations; one (ASA EG) contains explicit guidance for both the individual and the organization where that individual might work; and one (ESSC CoP) has guidance that could be useful to both the individual and the organization.

*Areas of strong alignment:* All four international guidances examined here were strongly aligned with the ASA EG. Alignment was particularly strong with UN FPOS, OECD GSP, and ISI DPE(denoted by green shading). (See summary Tables 1, 3, and 4.)

We note a somewhat weaker - indirect- alignment between the UN FPOS and OECD GSP, as denoted by the parenthesis. (See summary Table 5b) Indirect alignment is concerning because, since the language is not a clear match but rather, is implicit, it might require considerable assessment to determine whether one or the other (or neither, or both) guidances pertain in any situation. Since practitioners are traditionally not trained in ethical statistical practice, no matter how carefully they are trained in statistical practice itself, this is a significant concern and could undercut the utility of any such document.

*Areas of potential tension:* We also note many areas of mixed alignment, which we did not anticipate. See Summary Tables 5a, 5c-5f. Although there were several instances of alignment when examining specific guidelines, these were met with several other instances of tension or potential conflict, or both alignment and potential tension. Among these, ESSC CoP guidance appeared to have the most complicated alignment patterns across all guidances examined here. This may be an opportunity to consider if the differences observed relate to differences in professional language, clarify language to communicate shared norms, and/or provide more detailed guidance to practitioners to enable balancing tensions across guidances. Importantly, the prevalence of indirect alignment (checks in parentheses) and the presence of conflicting alignment with tension in the same cell for tables with ESSC CoP and OECD GSP and ISI DPEsignal a high likelihood for confusion or conflict.

*Guidance gap:* In a few cases, we observed apparent gaps where guidelines neither aligned nor departed from each other. See for example, guidelines relating to *a clear legal framework and rights to administrative data access*, which are supported directly or indirectly across all guidances examined here but are absent from ISI DPE. Overall, the gaps in guidance were not as concerning as they might be if there was no ethical practice standard with which any practitioner can assure they are using statistical practices ethically, irrespective of whether they are in, outside of, or working on teams with practitioners from national statistical offices.

*Discussion:* Overall, we observed less alignment and more tensions across guidances than we anticipated. Alignment was strongest with the ASA EG; weakest with the ESSC CoP, and gaps in alignment were most common in the ISI DPE.

## E. Discussion

In this series of analyses, we explored the extent to which international guidance for professional statistical practice align with the ASA Ethical Guidelines for Statistical Practice. We examined four key sources of international guidance in the professional conduct of statistical practice. Each of these guidances was developed for somewhat different audiences. Of course, the ASA EG are ethical guidelines for any and all statistical practices - without regard for the environment, job title, or preparation/degree of the practitioner. They contain specific guidance for individual practitioners, the organizations that employ individual practitioners, and those who lead, supervise, and mentor other practitioners. The UN FPOS, ESSC CoP, and OECD GSP are not ethical guidelines, but rather guidelines for "professional" statistical practice. All of these mention "professional independence" or "professional standards" for practice, and as such, it is understood that they are implicitly depending on ethical practice standards like the ASA EG as guidance for the individual practitioners employed by organizations that utilize these specific guidances. The ISI DPEis also not a set of guidelines but rather aspirational statements intended to facilitate communication and promote shared norms of professional conduct. Both the ASA and ISI are actively engaged in promoting growth in the awareness of, and compliance with, the ASA EGs as an ethical practice standard for statistics and data science.

The audiences for the UN FPOS, ESSC CoP, and OECD GSP substantially overlap geographically; coordinate institutionally; and comprise the statistics profession as a whole, albeit for a highly targeted domain (official statistics). These guidances were developed initially around the same time, and are routinely reviewed for possible revision to remain current. It is therefore reasonable to anticipate - and have observed - general alignment, particularly among international guidances.

We found very strong alignment of the ASA EG with the UN FPOS, the OECD GSP, and the ISI DPE. We also found strong alignment with the ASA EG and ESSC CoP. Our analyses did not show the extent -beyond what these international guidances articulate -of support for ethical statistical practices to the extent that these are reflected in the ASA EG.

We also examined alignment among the international guidances themselves. We were surprised to observe weaker alignment (UN FPOS and OECD GSP) and mixed alignment with tensions, particularly with ESSC CoP and the tendency for both alignment and tension among individual elements when examined against another international guidance. As might be expected, alignment was strongest with regard to norms of professional environment and conduct–particularly with regard to independence, integrity, and trust. When exploring specific areas of potential tension, it seemed that potential tensions were more common with regard to professional output. As noted in Park and Tractenberg (2023), tensions in balancing relevance, timeliness, accuracy, and objectivity are well-known to professional statisticians, and are not surprising to observe here.

Nonetheless, we also observed patterns of tension where guidelines explicitly call for cooperation and collaboration (such as ESSC CoP and OECD GSP). Adhering to these two principles in particular can apply peer pressures that may be positive (favoring data quality, for example) or negative (jeopardizing objectivity, for example). There also were some apparent gaps in coverage when examining alignment, such as principles relating to a clear legal framework and access to administrative data–absent from ISI DPE. When the ASA EG elements were aligned with "clear legal *and institutional* framework" (OECD GSP 1), ASA EG alignment was considerable (i.e., Table 3 shows alignment with ASA EG Principles A-E, G and H). This stems from the consideration of institutional strengths in the ASA EGs that the ISI, being aspirational rather than practical guidance, does not include. While legal, possibly more than institutional, frameworks are particularly important to national statistical offices (and data users) because they can enable the integrity and quality of official statistics, the omission from the ISI DPEis notable.

## F. Conclusions

Standards have a powerful role in the statistical profession. They are a means to communicate and recognize achievement of expectations of excellence, and thereby the public's trust. We teach students, and recognize professionals according to adherence to these standards. We consult standards at times of significant scientific advancement or other social change to provide guidance.

Standards are especially important when working with cross-disciplinary teams and multinational bodies. When clear, they support communication, conduct, and goals. When inconsistent or incomplete, standards can confuse, and cause tension or even apathy, and therefore, affect professional practice and scientific growth. Importantly, to our knowledge, this is the first investigation using an independently derived and comprehensive ethical practice standard for statistics and data science to explore how a practitioner who operates at the intersection of any of these guidance documents might encounter challenges when trying to follow one or more of these documents. Here, we examined alignment of four key international guidance documents with the ASA EG. Using a form of structured content analysis (see also Park and Tractenberg, 2023), we found unexpected results.

Across all guidelines examined, alignment was strongest between any guidance and the ASA EG. This supports the choice of the ASA EG as the ethical practice standard to utilize in these analyses. Alignment was mixed with potential for tension among all pairs of international guidance examined with the exception of UN FPOS and OECD GSP, where (somewhat weaker) alignment occurred. A common area of potential tension occurred, paradoxically, where guidelines call for cooperation and collaboration, such as in ESSC CoP and OECD GSP; such pressures can encourage or discourage independence, scientific methodology, and the output of quality statistics.

The individual practitioner is likely to find it challenging to adhere to guidance that appears conflicting (or is silent on some topics). The potential misalignments we identified suggest a scenario where a practitioner who follows one guidance document (UN, OECD, ESSC) will need to specifically address how s/he does not violate the other guidance, and/or the ASA EGs themselves. This paper makes these challenges explicit and describes how conflicts might arise.

How might these conflicts be avoided? In some cases, these tensions may reflect differences in priority, rather than conflict per se across guidance documents. One potential solution is for stewards of the guidances examined here to review and possibly clarify language where it appears to conflict. The ASA EG can be a resource for clarifying language, helping a practitioner to navigate situations when either a document itself is contradictory or otherwise does not provide information about how the practitioner should prioritize obligations.

Given the strength of alignment observed, and the scope and detail of practices described, the ASA EG can support achievement of the standards of the UN FPOS, ESSC CoP, OECD GSP, and ISI DPE. It can serve as a unifying ethical practice standard for statistics and data science, especially useful for multinational collaboration but definitely not limited to official statistics. We believe the detailed practices described in the ASA EG are strongly supportive of achieving cooperation and collaboration, in promoting the public trust in official statistics and in the profession and practice of statistics and data science.

### References

American Statistical Association (ASA), (2018; revised 2022). *ASA Ethical Guidelines for Statistical Practice-revised*, downloaded from https://www.amstat.org/ASA/Your-Career/Ethical-Guidelines-for-Statistical-Practice.aspx on 1 February 2022.

Campbell DT. (1975). "Degrees of Freedom" and the Case Study. *Comparative Political Studies*, 8:178–193.

Economic and Social Council, Statistical Commission and Economic Commission for Europe Conference of European Statisticians. *Book on The History of The CES: Chapter 8: The Fundamental Principles of Official Statistics: The Breakthrough of a New Era*. CES Working Paper 2 2002, Fiftieth plenary session of 2002, CES/2002/WP.2 (3 June 2002) https://unece.org/fileadmin/DAM/stats/documents/ece/ces/2002/wp.2.e.pdf

Committee on Statistics and Statistical Policy, Organisation for Economic Co-operation and Development, Recommendation of the Council on Good Statistical Practice. C/M(2019)7 1394th SESSION, agenda item 55 C(2019)28 (20 February 2019 adopted 13 March 2019). https://one.oecd.org/document/C(2019)28/en/pdf

Conference of European Statisticians. (1991). *The Fundamental Principles of Official Statistics in the Region of the ECE*. CES Res 702, 39th session, CES/702 (adopted 19 March 1991).

Economic Commission for Europe. (1992). *The Fundamental Principles of Official Statistics in the Region of the ECE.* C Res 47, 47th session, Decision C(47) (adopted 15 April 1992).

Economic and Social Council. (2013). *Fundamental Principles of Official Statistics.* ECOSOC Res 2013/21, Substantive session of 2013, Agenda item 13 (c) E/RES/2013/21 (28 October 2013 adopted 24 July). https://unstats.un.org/unsd/dnss/gp/FP-Rev2013-E.pdf

Eurostat. (2017). *European Statistics Code of Practice.* European Statistical System Committee EU Regulation 223/2009, Article 11 EU/223/2009 (Revision adopted 16th November 2017). http://ec.europa.eu/eurostat/documents/3859598/5921861/KS-32-11-955-EN.PDF

General Assembly. (2014). *Fundamental Principles of Official Statistics.* GA Res 68/261, Sixty-eighth session, agenda item 9 A/68/21 (3 March 2014 adopted 29 January 2014). https://unstats.un.org/unsd/dnss/gp/FP-New-E.pdf

European Union. (2018). *General Data Protection Rule.* Downloaded from https://gdpr.eu/tag/gdpr/

Gillikin J, Kopolow A & Schrimmer K. (2017). *Principles for the Development of a Professional Code of Ethics* [white paper]. National Association for Healthcare Quality. *SocArXiv* https://osf.io/preprints/socarxiv/dt5kr/

Hogan H & Steffey D. (2014). Professional Ethics for Statisticians: An Organizational History. *Proceedings of the 2014 Joint Statistical Meetings, Boston, MA.* Pp. 1397 -1404.

International Statistical Institute. (2010/2023). *Declaration on Professional Ethics.* Downloaded from https://www.isi-web.org/isi-declaration-professional-ethics-0

Nieswiadomy RM & Bailey C. (2018). *Foundations of Nursing Research*, 7th Edition. New York, NY: Pearson.

Park J & Tractenberg RE. (in press-2023). How do ASA Ethical Guidelines for Statistical Practice Support U.S. Guidelines for Official Statistics? In, H. Doosti, (Ed.). *Ethical Statistics.* Cambridge, UK: Ethics International Press. Preprint available at: *StatArXiv* http://arxiv.org/abs/2309.07180

Rios CM, Golde CM, Tractenberg RE. (2019). The Preparation of Stewards with the Mastery Rubric for Stewardship: Re-Envisioning the Formation of Scholars and Practitioners. E*ducation Sciences 9(4), 292;* https://doi.org/10.3390/educsci9040292

Tractenberg RE. (2019). Degrees of Freedom Analysis in educational research and decision-making: Leveraging qualitative data to promote excellence in bioinformatics training and education. *Briefings in Bioinformatics* 20(2): 416–425.

Tractenberg RE. (2020, 20 February). Concordance of professional ethical practice standards for the domain of Data Science: A white paper. Published in the *Open Archive of the Social Sciences (SocArXiv)*, DOI 10.31235/osf.io/p7rj2

Tractenberg RE. (2022-A). *Ethical Reasoning for a Data-Centered World.* Cambridge, UK: Ethics International Press.

Tractenberg RE. (2022-B). *Ethical Practice in Statistics and Data Science.* Cambridge, UK: Ethics International Press.

Tractenberg RE. (2023, May 6). Degrees of Freedom Analysis: A mixed method for theory building, decision making, and prediction. Published in the *Open Archive of the Social Sciences (SocArXiv)*, osf.io/preprints/socarxiv/r5a7z

Tractenberg RE & Gordon M. (2017). Supporting evidence-informed teaching in biomedical and health professions education through knowledge translation: An inter-disciplinary literature review. *Teaching and Learning in Medicine* 29 (3): 268-279. http://dx.doi.org/10.1080/10401334.2017.1287572

Wilson EJ & Wilson DT. (1988). "Degrees of Freedom" in Case Research of Behavioral Theories of Group Buying. in MJ Houston (Ed.). *Advances in Consumer Research, Volume 15*. Provo, UT: Association for Consumer Research. Pp. 587-594.

Wilson EJ & Vlosky RP. (1997). Partnering Relationship Activities: Building Theory from Case Study Research. *Journal of Business Research*, 39:59–70.

Wilson EJ & Woodside AG. (1999). Degrees-of-Freedom analysis of case data in business marketing research. *Industrial Marketing Management,* 28:215-229.

Woodside AG, & Wilson EJ. (2003). Case study research for theory-building. *Journal of Business & Industrial Marketing*, 18:493–508.

Woodside AG. (2010). *Case Study Research: Theory, Methods and Practice*. Bangles, UK: Emerald Groups.

**Annex**

1 ASA Ethical Guidelines

2 Detail Tables

### *Annex 1. ASA Ethical Guidelines for Statistical Practice, 2022*

**Ethical Guidelines for Statistical Practice**
***Prepared by the Committee on Professional Ethics***
***of the American Statistical Association***
***February 2022***

## PURPOSE OF THE GUIDELINES:

The American Statistical Association's Ethical Guidelines for Statistical Practice are intended to help statistical practitioners make decisions ethically. In these Guidelines, "statistical practice" includes activities such as: designing the collection of, summarizing, processing, analyzing, interpreting, or presenting, data; as well as model or algorithm development and deployment. Throughout these Guidelines, the term "statistical practitioner" includes all those who engage in statistical practice, regardless of job title, profession, level, or field of degree. The Guidelines are intended for individuals, but these principles are also relevant to organizations that engage in statistical practice.

The Ethical Guidelines aim to promote accountability by informing those who rely on any aspects of statistical practice of the standards that they should expect. Society benefits from informed judgments supported by ethical statistical practice. All statistical practitioners are expected to follow these Guidelines and to encourage others to do the same.

In some situations, Guideline principles may require balancing of competing interests. If an unexpected ethical challenge arises, the ethical practitioner seeks guidance, not exceptions, in the Guidelines. To justify unethical behaviors, or to exploit gaps in the Guidelines, is unprofessional, and inconsistent with these Guidelines.

### PRINCIPLE A: Professional Integrity and Accountability

Professional integrity and accountability require taking responsibility for one's work. Ethical statistical practice supports valid and prudent decision making with appropriate methodology. The ethical statistical practitioner represents their capabilities and activities honestly, and treats others with respect.

**The ethical statistical practitioner:**

1. Takes responsibility for evaluating potential tasks, assessing whether they have (or can attain) sufficient competence to execute each task, and that the work and timeline are feasible. Does not solicit or deliver work for which they are not qualified, or that they would not be willing to have peer reviewed.
2. Uses methodology and data that are valid, relevant, and appropriate, without favoritism or prejudice, and in a manner intended to produce valid, interpretable, and reproducible results.
3. Does not knowingly conduct statistical practices that exploit vulnerable populations or create or perpetuate unfair outcomes.
4. Opposes efforts to predetermine or influence the results of statistical practices, and resists pressure to selectively interpret data.
5. Accepts full responsibility for their own work; does not take credit for the work of others; and gives credit to those who contribute. Respects and acknowledges the intellectual property of others.
6. Strives to follow, and encourages all collaborators to follow, an established protocol for authorship. Advocates for recognition commensurate with each person's contribution to the work. Recognizes that inclusion as an author does imply, while acknowledgement may imply, endorsement of the work.
7. Discloses tensions of interest, financial and otherwise, and manages or resolves them according to established policies, regulations, and laws.
8. Promotes the dignity and fair treatment of all people. Neither engages in nor condones discrimination based on personal characteristics. Respects personal boundaries in interactions and avoids harassment including sexual harassment, bullying, and other abuses of power or authority.
9. Takes appropriate action when aware of deviations from these Guidelines by others.
10. Acquires and maintains competence through upgrading of skills as needed to maintain a high standard of practice.
11. Follows applicable policies, regulations, and laws relating to their professional work, unless there is a compelling ethical justification to do otherwise.
12. Upholds, respects, and promotes these Guidelines. Those who teach, train, or mentor in statistical practice have a special obligation to promote behavior that is consistent with these Guidelines.

## PRINCIPLE B: Integrity of Data and Methods

The ethical statistical practitioner seeks to understand and mitigate known or suspected limitations, defects, or biases in the data or methods and communicates potential impacts on the interpretation, conclusions, recommendations, decisions, or other results of statistical practices.

**The ethical statistical practitioner:**

1.   Communicates data sources and fitness for use, including data generation and collection processes and known biases. Discloses and manages any tensions of interest relating to the data sources. Communicates data processing and transformation procedures, including missing data handling.

2.   Is transparent about assumptions made in the execution and interpretation of statistical practices including methods used, limitations, possible sources of error, and algorithmic biases. Conveys results or applications of statistical practices in ways that are honest and meaningful.

3.   Communicates the stated purpose and the intended use of statistical practices. Is transparent regarding a priori versus post hoc objectives and planned versus unplanned statistical practices. Discloses when multiple comparisons are conducted, and any relevant adjustments.

4.   Meets obligations to share the data used in the statistical practices, for example, for peer review and replication, as allowable. Respects expectations of data contributors when using or sharing data. Exercises due caution to protect proprietary and confidential data, including all data that might inappropriately harm data subjects.

5.   Strives to promptly correct substantive errors discovered after publication or implementation. As appropriate, disseminates the correction publicly and/or to others relying on the results.

6.   For models and algorithms designed to inform or implement decisions repeatedly, develops and/or implements plans to validate assumptions and assess performance over time, as needed. Considers criteria and mitigation plans for model or algorithm failure and retirement.

7.   Explores and describes the effect of variation in human characteristics and groups on statistical practice when feasible and relevant.

## PRINCIPLE C: Responsibilities to Stakeholders

Those who fund, contribute to, use, or are affected by statistical practices are considered stakeholders. The ethical statistical practitioner respects the interests of stakeholders while practicing in compliance with these Guidelines.

**The ethical statistical practitioner:**

1. Seeks to establish what stakeholders hope to obtain from any specific project. Strives to obtain sufficient subject-matter knowledge to conduct meaningful and relevant statistical practice.
2. Regardless of personal or institutional interests or external pressures, does not use statistical practices to mislead any stakeholder.
3. Uses practices appropriate to exploratory and confirmatory phases of a project, differentiating findings from each so the stakeholders can understand and apply the results.
4. Informs stakeholders of the potential limitations on use and re-use of statistical practices in different contexts and offers guidance and alternatives, where appropriate, about scope, cost, and precision considerations that affect the utility of the statistical practice.
5. Explains any expected adverse consequences from failing to follow through on an agreed-upon sampling or analytic plan.
6. Strives to make new methodological knowledge widely available to provide benefits to society at large. Presents relevant findings, when possible, to advance public knowledge.
7. Understands and conforms to confidentiality requirements for data collection, release, and dissemination and any restrictions on its use established by the data provider (to the extent legally required). Protects the use and disclosure of data accordingly. Safeguards privileged information of the employer, client, or funder.
8. Prioritizes both scientific integrity and the principles outlined in these Guidelines when interests are in conflict.

## PRINCIPLE D: Responsibilities to Research Subjects, Data Subjects, or those directly affected by statistical practices

The ethical statistical practitioner does not misuse or condone the misuse of data. They protect and respect the rights and interests of human and animal subjects. These responsibilities extend to those who will be directly affected by statistical practices.

**The ethical statistical practitioner:**

1. Keeps informed about and adheres to applicable rules, approvals, and guidelines for the protection and welfare of human and animal subjects. Knows when work requires ethical review and oversight.[1]
2. Makes informed recommendations for sample size and statistical practice methodology in order to avoid the use of excessive or inadequate numbers of subjects and excessive risk to subjects
3. For animal studies, seeks to leverage statistical practice to reduce the number of animals used, refine experiments to increase the humane treatment of animals, and replace animal use where possible.

4.  Protects people's privacy and the confidentiality of data concerning them, whether obtained from the individuals directly, other persons, or existing records. Knows and adheres to applicable rules, consents, and guidelines to protect private information.

5.  Uses data only as permitted by data subjects' consent when applicable or considering their interests and welfare when consent is not required. This includes primary and secondary uses, use of repurposed data, sharing data, and linking data with additional data sets.

6.  Considers the impact of statistical practice on society, groups, and individuals. Recognizes that statistical practice could adversely affect groups or the public perception of groups, including marginalized groups. Considers approaches to minimize negative impacts in applications or in framing results in reporting.

7.  Refrains from collecting or using more data than is necessary. Uses confidential information only when permitted and only to the extent necessary. Seeks to minimize the risk of re-identification when sharing de-identified data or results where there is an expectation of confidentiality. Explains any impact of de-identification on accuracy of results.

8.  To maximize contributions of data subjects, considers how best to use available data sources for exploration, training, testing, validation, or replication as needed for the application. The ethical statistical practitioner appropriately discloses how the data is used for these purposes and any limitations.

9.  Knows the legal limitations on privacy and confidentiality assurances and does not over-promise or assume legal privacy and confidentiality protections where they may not apply.

10. Understands the provenance of the data, including origins, revisions, and any restrictions on usage, and fitness for use prior to conducting statistical practices.

11. Does not conduct statistical practice that could reasonably be interpreted by subjects as sanctioning a violation of their rights. Seeks to use statistical practices to promote the just and impartial treatment of all individuals.

### PRINCIPLE E: Responsibilities to members of multidisciplinary teams

Statistical practice is often conducted in teams made up of professionals with different professional standards. The statistical practitioner must know how to work ethically in this environment.

**The ethical statistical practitioner:**

1.  Recognizes and respects that other professions may have different ethical standards and obligations. Dissonance in ethics may still arise even if all members feel that they are working towards the same goal. It is essential to have a respectful exchange of views.

2.  Prioritizes these Guidelines for the conduct of statistical practice in cases where ethical guidelines conflict.

3.  Ensures that all communications regarding statistical practices are consistent with these Guidelines. Promotes transparency in all statistical practices.

4.  Avoids compromising validity for expediency. Regardless of pressure on or within the team, does not use inappropriate statistical practices.

**PRINCIPLE F: Responsibilities to Fellow Statistical Practitioners and the Profession**

Statistical practices occur in a wide range of contexts. Irrespective of job title and training, those who practice statistics have a responsibility to treat statistical practitioners, and the profession, with respect. Responsibilities to other practitioners and the profession include honest communication and engagement that can strengthen the work of others and the profession.

**The ethical statistical practitioner:**

1.  Recognizes that statistical practitioners may have different expertise and experiences, which may lead to divergent judgments about statistical practices and results. Constructive discourse with mutual respect focuses on scientific principles and methodology and not personal attributes.
2.  Helps strengthen, and does not undermine, the work of others through appropriate peer review or consultation. Provides feedback or advice that is impartial, constructive, and objective.
3.  Takes full responsibility for their contributions as instructors, mentors, and supervisors of statistical practice by ensuring their best teaching and advising -- regardless of an academic or non-academic setting -- to ensure that developing practitioners are guided effectively as they learn and grow in their careers.
4.  Promotes reproducibility and replication, whether results are "significant" or not, by sharing data, methods, and documentation to the extent possible.
5.  Serves as an ambassador for statistical practice by promoting thoughtful choices about data acquisition, analytic procedures, and data structures among non-practitioners and students. Instills appreciation for the concepts and methods of statistical practice.

**PRINCIPLE G: Responsibilities of Leaders, Supervisors, and Mentors in Statistical Practice**

Statistical practitioners leading, supervising, and/or mentoring people in statistical practice have specific obligations to follow and promote these Ethical Guidelines. Their support for – and insistence on – ethical statistical practice are essential for the integrity of the practice and profession of statistics as well as the practitioners themselves.

**Those leading, supervising, or mentoring statistical practitioners are expected to**:

1.  Ensure appropriate statistical practice that is consistent with these Guidelines. Protect the statistical practitioners who comply with these Guidelines, and advocate for a culture that supports ethical statistical practice.
2.  Promote a respectful, safe, and productive work environment. Encourage constructive engagement to improve statistical practice.
3.  Identify and/or create opportunities for team members/mentees to develop professionally and maintain their proficiency.
4.  Advocate for appropriate, timely, inclusion and participation of statistical practitioners as contributors/collaborators. Promote appropriate recognition of the contributions of statistical practitioners, including authorship if applicable.

5. Establish a culture that values validation of assumptions, and assessment of model/algorithm performance over time and across relevant subgroups, as needed. Communicate with relevant stakeholders regarding model or algorithm maintenance, failure, or actual or proposed modifications.

## PRINCIPLE H: Responsibilities Regarding Potential Misconduct

The ethical statistical practitioner understands that questions may arise concerning potential misconduct related to statistical, scientific, or professional practice. At times, a practitioner may accuse someone of misconduct, or be accused by others. At other times, a practitioner may be involved in the investigation of others' behavior. Allegations of misconduct may arise within different institutions with different standards and potentially different outcomes. The elements that follow relate specifically to allegations of statistical, scientific, and professional misconduct.

**The ethical statistical practitioner:**

1. Knows the definitions of, and procedures relating to, misconduct in their institutional setting. Seeks to clarify facts and intent before alleging misconduct by others. Recognizes that differences of opinion and honest error do not constitute unethical behavior.
2. Avoids condoning or appearing to condone statistical, scientific, or professional misconduct. Encourages other practitioners to avoid misconduct or the appearance of misconduct.
3. Does not make allegations that are poorly founded, or intended to intimidate. Recognizes such allegations as potential ethics violations.
4. Lodges complaints of misconduct discreetly and to the relevant institutional body. Does not act on allegations of misconduct without appropriate institutional referral, including those allegations originating from social media accounts or email listservs.
5. Insists upon a transparent and fair process to adjudicate claims of misconduct. Maintains confidentiality when participating in an investigation. Discloses the investigation results honestly to appropriate parties and stakeholders once they are available.
6. Refuses to publicly question or discredit the reputation of a person based on a specific accusation of misconduct while due process continues to unfold.
7. Following an investigation of misconduct, supports the efforts of all parties involved to resume their careers in as normal a manner as possible, consistent with the outcome of the investigation.
8. Avoids, and acts to discourage, retaliation against or damage to the employability of those who responsibly call attention to possible misconduct.

## APPENDIX
## Responsibilities of organizations/institutions

Whenever organizations and institutions design the collection of, summarize, process, analyze, interpret, or present, data; or develop and/or deploy models or algorithms, they have responsibilities to use statistical practice in ways that are consistent with these Guidelines, as well as promote ethical statistical practice.

**Organizations and institutions engage in, and promote, ethical statistical practice by**:

1. Expecting and encouraging all employees and vendors who conduct statistical practice to adhere to these Guidelines. Promoting a workplace where the ethical practitioner may apply the Guidelines without being intimidated or coerced. Protecting statistical practitioners who comply with these Guidelines.
2. Engaging competent personnel to conduct statistical practice, and promote a productive work environment.
3. Promoting the professional development and maintenance of proficiency for employed statistical practitioners.
4. Supporting statistical practice that is objective and transparent. Not allowing organizational objectives or expectations to encourage unethical statistical practice by its employees.
5. Recognizing that the inclusion of statistical practitioners as authors, or acknowledgement of their contributions to projects or publications, requires their explicit permission because it may imply endorsement of the work.
6. Avoiding statistical practices that exploit vulnerable populations or create or perpetuate discrimination or unjust outcomes. Considering both scientific validity and impact on societal and human well-being that results from the organization's statistical practice.
7. Using professional qualifications and contributions as the basis for decisions regarding statistical practitioners' hiring, firing, promotion, work assignments, publications and presentations, candidacy for offices and awards, funding or approval of research, and other professional matters.

**Those in leadership, supervisory, or managerial positions who oversee statistical practitioners promote ethical statistical practice by following Principle G and:**

8. Recognizing that it is contrary to these Guidelines to report or follow only those results that conform to expectations without explicitly acknowledging competing findings and the basis for choices regarding which results to report, use, and/or cite.
9. Recognizing that the results of valid statistical studies cannot be guaranteed to conform to the expectations or desires of those commissioning the study or employing/supervising the statistical practitioner(s).
10. Objectively, accurately, and efficiently communicating a team's or practitioners' statistical work throughout the organization.
11. In cases where ethical issues are raised, representing them fairly within the organization's leadership team.
12. Managing resources and organizational strategy to direct teams of statistical practitioners along the most productive lines in light of the ethical standards contained in these Guidelines.

---

[1] Examples of ethical review and oversight include an Institutional Review Board (IRB), an Institutional Animal Care and Use Committee (IACUC), or a compliance assessment.

**Annex 2: Detailed Tables**

Appendix Table 1: ASA Ethical Guidelines (2022) with UN FPOS (2014)

| ASA Ethical Guidelines / UN Fundamental Principles of Official Statistics | A Professional Integrity and Accountability | B Integrity of Data and Methods | C Stakeholders | D Research Subjects/Data Subjects and Those Affected by Statistical Practices | E Interdisciplinary Team Members | F Other Practitioners/ Profession | G Leader/ Supervisor/ Mentor and Appendix (APP) | H Allegations of Potential Misconduct |
|---|---|---|---|---|---|---|---|---|
| U1. Official statistics that meet the test of practical utility are to be compiled and made available publicly. | A2 | B, B1, B3, B6 | C1, C4 | D2, D6, D8, D10 | E4 | | G1, G5; APP 1, 2, 4,9 | H2 |
| U2. Statistical agencies choose methods and procedures for the collection, processing, storage and presentation of statistical data based on strictly professional considerations, including scientific principles and professional ethics | A, A2, A3, A4, A11, A12 | B, B1, B2, B3 | C, C1, C2, C4, C8 | D1, D2, D4, D5, D6, D8, D10, D11 | E2, E3, E4 | F4 | G1, G2, G5; APP 1, 2, 4, 6, 8, 9, 11 | H1, H2 |
| U3. Statistical agencies are to present information according to scientific standards on the sources, methods and procedures of the statistics. | A, A2, A3, A4, A11, A12 | B, B1, B2, B3 | C2, C8 | D1, D2, D4, D5, D6, D8, D10, D11 | E4 | F4 | G1, G5; APP 1, 2, 4, 6, 8, 9, 11 | H2 |
| U4. The statistical agencies are entitled to comment on erroneous interpretation and misuse of statistics. | A2, A4, A9 | B5a | C2 | D11 | E4 | | APP 8, 9, 10 | H2 |
| U5. Statistical agencies choose data sources with regard to quality, timeliness, costs and the burden on respondents. | (~A2) | B1 | C4 | D, D1, D4, D10, D11 | E4 | | APP 6 | H2 |
| U6. Confidential data collection used exclusively for statistical purposes. | | | C7 | D, D1, D4, D5, D9, D10 | | | | |
| U7. Public laws, regulations and measures under which the statistical systems operate. | A7, A11 | B4 | C7, C8 | D1, D4, D9, D10, D11 | E2 | | G1; APP 1, 4, 8 | H, H1, H4 |
| U8. Coordination among statistical agencies within countries | A4, A9, A11 | B1, B2, B3 | C1, C2, C4, C8 | D10, D11 | E2, E4 | F5 | G1, G5; APP 1, 8, 9, 10, 11 | |
| U9. The consistent use by statistical agencies in each country of international concepts, classifications and methods | A2 | B2, B5 | C2, C8 | D10 | E3 | F4 | G1, G5; APP 1, 4, 9, 10, 11 | |
| U10. Statistical agencies cooperate bilaterally and multilaterally | A2 | B2, B5 | C2, C8 | D10 | E3 | F4 | G1, G5; APP 1, 4, 9, 10, 11 | |

a *Obliged*, not entitled.

Appendix Table 2: ASA Ethical Guidelines (2022) with ESSC CoP (2017)

| ASA Ethical Guidelines / ESSC Code of Statistical Practice | A Professional Integrity and Accountability | B Integrity of Data and Methods | C Stakeholders | D Research Subjects/Data Subjects and Those Affected by Statistical Practices | E Interdisciplinary Team Members | F Other Practitioners/Profession | G Leader/ Supervisor/ Mentor and Appendix (APP) | H Allegations of Potential Misconduct |
|---|---|---|---|---|---|---|---|---|
| Institutional Environment | | | | | | | | |
| E1. Professional Independence | A, A1, A2, (A3), A4, A5, A9, A11, A12 | B, B2, B3, B5, B6 | C1, C2, C8 | D1, D5, D6, D9, D10, D11 | E2, E3, E4 | F4 (F5) | G1, G2, G5; APP 1, 2, 4, 6, 7, 8, 9, 10 | (H2) |
| E1bis. Coordination and Cooperation | A, A1, A2, A3, A4, A5, (A7) | B, B1, B2, B3, B4, B5 | C, C1, C2, C4 | D, D1, D4, D5, D6, D7, D10, D11 | (E1),~E4 | (F1) | G1, G2, G5; APP 11 | H2 |
| E2. Mandate for Data Collection and Access to Data | A2,~A3 | B4 | C7 | D, D4, D5, (D7, D8, D9). D10, D11 | E4 | | | |
| E3. Adequacy of Resources | (A1) | | | | | | APP 12 | |
| E4. Commitment to Quality | A, A1, A2, A3, A4, A10, A12 | B, B5, B6 | C1, C2 | D, D5, D6, D11 | E4 | F2 | G, G1, G2, G5; APP 1, 2, 3, 4, 6, 7, 8, 9, 10,12 | H2 |
| E5. Statistical Confidentiality and Data Protection | A11 | B4 | C7 | D, D4, D5, D7, D9, D10 | E4 | | | |
| Statistical Processes | | | | | | | | |
| E6. Impartiality and Objectivity | A1, A2, A4, A5, A7 | B, B1, B2, B3, B5, B6 | C2, C8 | | E2, E4 | F4 | G1,G5; APP 1, 2, 4, 8, 9, 10 | |
| E7. Sound Methodology | A2, A4 | B2, B3 | C1, C2 | D2, D10 | E4 | | G1 (G5); APP 1, 2, 4, 8, 9 | H2 |
| E8. Appropriate Statistical Procedures | A2, A4 | B6 (B7) | C2, C3 | D, D5, D8, D10, D11 | E4 | | G1, G5; APP 1, 2, (3), 4, 6, 8, 9, 10) | H2 |
| E9. No Undue Respondent Burden | | | | (D2, D7) | | | | |
| E10. Cost Effectiveness | | | | (D2, D3, D7, D8) | | | APP (12) | |
| Statistical Output | | | | | | | | |
| E11. Relevance | A2, ~A3, ~A4 | B, (B3), ~B1, ~B2 | C1, ~C2 | (~D, ~D11) | E4 | | G1, G5 | |
| E12. Accuracy and Reliability | A2, A4 | B, B1, B2, B3, B5, B6 | C2 | | E4 | | G1, G5; APP 1, 2, 4, 8, 9, 10 | |
| E13. Timeliness and Punctuality | ~A2, ~A4 | | ~C2, ~C8 | ~D10 | E4 | | ~G1, ~G5; APP 1, 2, 8, 9, 10 | |
| E14. Coherence and Comparability | A2, A4, ~A2 | (B, B5),~B6 | C2 | (~D11) | ~E4 | F4 | ~G1, ~G5; APP 1, 2, ~8, ~9, ~10 | (H2) |
| E15. Accessibility and Clarity | A,~A4 | B1, B2, B3 | C2, C4 | D8 | E3 | F2, F4 | G5; APP 4,10, 11 | H1 |

Appendix Table 3: ASA Ethical Guidelines (2022) with OECD GSP (2014)

| ASA Ethical Guidelines / OECD Good Statistical Practice | A Professional Integrity and Accountability | B Integrity of Data and Methods | C Stakeholders | D Research Subjects/Data Subjects and Those Affected by Statistical Practices | E Interdisciplinary Team Members | F Other Practitioners/Profession | G Leader/ Supervisor/ Mentor and Appendix (APP) | H Allegations of Potential Misconduct |
|---|---|---|---|---|---|---|---|---|
| O1. A clear legal and institutional framework | A7, A11 | B4 | C7, C8 | D1, D4, D9, D10, D11 | E2 | | G1; APP 1, 4, 8, 9, 10 | H, H1, H4 |
| O2. Professional independence | A, A2, A3, A4, A11, A12 | B, B1, B2, B3, B5 | C2, C8 | D1, D2, D4, D5, D6, D8, D10, D11 | E4 | F4 | G1, G2, G5; APP1, 2, 4, 6, 8, 9, 11 | H2 |
| O3. Adequacy of resources | (A1) | | | | | | APP 12 | |
| O4. Protection of privacy | | | C7 | D, D1, D4, D5, D9, D10 | | | | |
| O5. The right to access administrative sources | A2, A3, A4 | | C4 | D, D1, D4, D10, D11 | E4 | | APP 6 | H2 |
| O6. Impartiality, objectivity and transparency | A, A2, A3, A4, A11, A12 | B, B1, B2, B3 | C, C1, C2, C4, C8 | D1, D2, D4, D5, D6, D8, D10, D11 | E2, E3, E4 | F4 | G1, G2, G5; APP1, 2, 4, 6, 8, 9, 11 | H1, H2 |
| O7. Sound methodology and professional standards | A, A2, A3, A4, A5, A7, A8 | B, B1, B2 | C2, C3, C8 | D, D2, D5, D10, D11 | E2, E4 | F4 | G1, G5; APP 1, 2, 4, 6, 8, 9, 10, 11 | H2 |
| O8. The quality of statistical outputs and processes | A2, A3, A4 | B, B1, B2, B3, B5, B6, B7 | C, C1, C2, C3, C4, C5, C8 | D, D2, D6, D10, D11 | E3 | F4 | G1, G5; APP 1, 2, 4, 6, 8, 9, 10, 11 | H2 |
| O9. User-friendly access and dissemination of data and metadata, and a commitment to respond to major misinterpretations of data by users | A2, A4, A9 | B5 | C2 | D11 | E4 | F4 | APP 4, 8, 9, 10, 11 | H2 |
| O10. Coordination of statistical activities | A4, A9, A11 | B1, B2, B3 | C1, C2, C4, C8 | D10, D11 | E2, E4 | F5 | G1, G5; APP 1, 8, 9, 10, 11 | |
| O11. International cooperation | A2 | B2, B5 | C2, C8 | D10 | E3 | F4 | G1, G5; APP 1, 4, 9, 10, 11 | |
| O12. Exploring new and alternative data sources and methods | A2, ~A3 | B4, B6, B7 | C3, C4, C6 | D8, ~D, ~D4, ~D5, ~D6, ~D7, ~D11 | | | | H2 |

Appendix Table 4. ASA Ethical Guidelines (2022) and ISI Statement of Ethics (2023)

| ASA Ethical Guidelines<br><br><br><br><br>ISI Statement of Ethics | A<br>Professional Integrity and Accountability | B<br>Integrity of Data and Methods | C<br>Stakeholders | D<br>Research Subjects /Data Subjects and Those Affected by Statistical Practices | E<br>Interdisciplinary Team Members | F<br>Other Practitioners/ Profession | G<br>Leader/ Supervisor/ Mentor and Appendix (APP) | H<br>Allegations of Potential Misconduct |
|---|---|---|---|---|---|---|---|---|
| **Shared Professional Values** | | | | | | | | |
| I1. Respect for data providers, communities, and others' work | (A2), A3, (A4), A5, (A8, A9) | (B4) | (C2), C5, C7 | D, D4, D5, (D6, D7), D9, D11 | | (F; F1) F2 | (G; G2); APP 6, 10 | |
| I2. Professionalism: responsibility; competence; and Informed Judgment | A;A1, A2, A5, A10 (A4, A7, A9, A11, ~A11, A12) | B; B1, B2, B3 (B5, B6) | C; C1, C2 (C6, C8) | | E2, E3, E4 | F; F2, F3, F4, (F5) | G; G1, G5; APP 1, 2 (10, 12) | (H2) |
| I3. Truthfulness and integrity: independence,; objectivity; transparency | A; A1, A2, A3, A4, A5 (A6) | B; B1, B2, B3, B5 | C; C1, C2, C4, C6, C8 | (D10) | E2, E3, E4 | (F2, F4) | (G; G1, G5); APP 1, (2), 4, 8, 9, 10, (11) | H; H2 |
| **Ethical Principles** | | | | | | | | |
| I1. Pursue objectivity by transparent methods and communication | A; A2, A4, A5 (A7, A9) | B; B1, B2, B3, B5 (B6) | C; C2, C4, C5, C6 (C8) | D; D8 (D10, D11) | E; E1, E2, E3, E4 | F; F4, F5 | G; G1, G2, G5; APP 1, 4, 8, 9, 10, 11 | H2 |
| I2. Clarify obligations and roles by area of competency | A; A1; (A3) A5; A6 (A10) | | C1 | D (D7) | E; E2, E3, E4 | F; (F3, F5) | G; (G1, G2, G4, G5) APP 1, (2), 5 (8, 9, 11) | H; H2 |
| I3. Assess Alternatives Impartially | (A4) | B; B1, B2, B3, B6 (B7) | (C2) C4, C5 | (D6, D7, D8) | (E3, E4) | F5 | (G5) APP (1, 4, 10) | |
| I4. Avoid Conflicting Interests | A7 | B1 | C; C2 | (D;D1) | E3 | F | (G1) APP 4, 8, 9 | (H2) |
| I5. Avoid Preempted Outcomes | A4, (A5, A7, A9, A12) | B; B2, B3 | C2 (C8) | (D10) | E2, E3, E4 | F4 | G;G1; APP 4, 8, 9 | H2 |
| I6. Guard Privileged Information | (A3) | (B4) | (C7) | D; D1, D4, D7, D9, D10, D11 | (E4) | | | (H2) |
| I7. Exhibit Professional Competence | A10 (A; A1) | | | (D1) | | (F2, F3) | (G) G3; APP 3 (11) | |
| I8. Maintaining Confidence in Statistics | (A; A1, A2, A4, A5) | B; B1, B2, B3, B5, B6 | C; C2, C3, C4, C8 | D (D6, D10) | E; E1, E2, E3, E4 | (F4; F5) | G; G1, G5; APP 1, 2, 4 (8, 9, 10, 11) | (H2) |
| I9. Exposing and Reviewing Methods and Findings | A; A1, A2, A4 (A12) | B; B1, B2, B3 (B5, B6) | C2, C3, C4, C6 (C8) | D8, (D6) | (E2, E3, E4) | F; (F2), F4 | G; G1, G2, G5; APP (1, 2, 8, 9, 10) | (H2) |

International Guidelines align with ASA Ethical Guidelines

| ASA Ethical Guidelines / ISI Statement of Ethics | A Professional Integrity and Accountability | B Integrity of Data and Methods | C Stakeholders | D Research Subjects /Data Subjects and Those Affected by Statistical Practices | E Interdisciplinary Team Members | F Other Practitioners/ Profession | G Leader/ Supervisor/ Mentor and Appendix (APP) | H Allegations of Potential Misconduct |
|---|---|---|---|---|---|---|---|---|
| I10. Communicating Ethical Principles | A9, A11, A12 | | (C8) | (D8, D11) | E; (E1), E2, E3, E4 | F; (F3, F5) | G; G1, G2, G5; APP 1, 3, 4, 11 | H2 |
| I11. Responsibility for the Integrity of the Discipline | A; A1, A2, A4 (A12) | B; B1, B2, B3,B5, B6 | C; C4, C8, C2, C3 | D; D1 (D5) D8, D11 | E; E2, E3, E4 | F; (F2) F4; F5 | G; G1, G2, G5; APP 1, 4, 8, 9, 10, (11) | H1, H2 |
| I12. Protecting the Interests of Subjects | (A3) | (B4) | C | D; D1, D2, D4, D5, D6, D7, D8, D9 (D10) | (E4) | | (G1; G5); APP (4, 5) | |

International Guidelines align with ASA Ethical Guidelines

Appendix Table 5a. UN Fundamental Principles of Official Statistics (2014) and ESSC European Code of Statistical Practice (2017)

| UN Fundamental Principles of Official Statistics<br><br>ESSC European Code of Practice | U1. Make data publicly available | U2. Collect using scientific methods | U3. Present using professional standards | U4. Comment on misuse | U5. Choose data considering quality, timeliness, cost and burden | U6. Use confidential data for statistical purposes only | U7. Operate the statistical system within a legal framework | U8. Coordinate across national statistical agencies | U9. Use international statistical classifications consistently | U10. Cooperate across national statistical systems |
|---|---|---|---|---|---|---|---|---|---|---|
| **Institutional Environment** | | | | | | | | | | |
| E1. Professional Independence | | x | | | | x | | x | | ~ |
| E1bis: Coordination and cooperation | | (~) | | (~) | | (~) | (~) | (~) | | x |
| E2. Mandate for Data Collection and Access to Data | (x) | | x | x | x | | (x) | | | x |
| E3. Adequacy of Resources | | | x | x | x | | | | | x |
| E4. Commitment to Quality | | x | | x | | x | x | x | x | |
| E5. Statistical Confidentiality and Data Protection | x | x | (x) | x | x | | x | x | | (~) |
| E6. Impartiality and Objectivity | | x | x | | x | x | x | x | | (~) |
| **Statistical Processes** | | | | | | | | | | |
| E7. Sound Methodology | | x | x | | | | | | | |
| E8. Appropriate Statistical Procedures | | x | x | x | | | | | | |
| E9. Non-excessive Burden on Respondents | | | | | x | | | | | |
| E10. Cost Effectiveness | | | | | x | | | | | |
| **Statistical Output** | | | | | | | | | | |
| E11. Relevance | | | x | | | | | | | |
| E12. Accuracy and Reliability | | x | x | (x) | | | | | | |
| E13. Timeliness and Punctuality | | | | | x | | | | | |
| E14. Coherence and Comparability | | | | | ~ | | | x | x | x |
| E15. Accessibility and Clarity | x | x | x | x | | | | x | ~ | x |

Appendix Table 5b: UN Fundamental Principles of Official Statistics (2014) and OECD Good Statistical Practice (2015)

| UN Fundamental Principles of Official Statistics / OECD Good Statistical Practice | U1. Make data publicly available | U2. Collect using scientific methods | U3. Present using professional standards | U4. Comment on misuse | U5. Choose data considering quality, timeliness, cost and burden | U6. Use confidential data for statistical purposes only | U7. Operate the statistical system within a legal framework | U8. Coordinate across national statistical agencies | U9. Use international statistical classifications consistently | U10. Cooperate across national statistical systems |
|---|---|---|---|---|---|---|---|---|---|---|
| O1. A clear legal and institutional framework | | | | | | | x | | | |
| O2. Professional independence | | (x) | | | x | | | ~x | ~x | ~x |
| O3. Adequacy of resources | | | | | | | (x) | | | |
| O4. Protection of privacy | | | | | | x | | | | |
| O5. The right to access administrative sources | x | | | | x | x | | (x) | (x) | (x) |
| O6. Impartiality, objectivity and transparency | x | x | x | x | | | | | | |
| O7. Sound methodology and professional standards | x | x | x | x | | | | | | |
| O8. Quality of statistical outputs and processes | | x | x | x | | | | | | |
| O9. User-friendly access and commitment to respond to user misinterpretation | x | | | x | | | | | | |
| O10. Coordination of statistical activities | | | | | | | | x | x | x |
| O11. International cooperation | | | | | | | | | x | x |
| O12. Explore new and alternative data sources/methods | x | x | x | x | | | | | | |

NB: UN 8, 9, 10 have the potential to undermine OECD2 (professional independence). The inconsistencies (misalignments) mean that those who follow one guidance document (UN or OECD) will need to specifically address how they do not violate the other.

# International Guidelines align with ASA Ethical Guidelines

Appendix Table 5c. UN Fundamental Principles of Official Statistics (2014) and ISI Statement of Ethics (2023)

| UN Fundamental Principles of Official Statistics / ISI Statement of Ethics | U1. Make data publicly available | U2. Collect using scientific methods | U3. Present using professional standards | U4. Comment on misuse | U5. Choose data considering quality, timeliness, cost and burden | U6. Use confidential data for statistical purposes only | U7. Operate the statistical system within a legal framework | U8. Coordinate across national statistical agencies | U9. Use international statistical classifications consistently | U10. Cooperate across national statistical systems |
|---|---|---|---|---|---|---|---|---|---|---|
| **Shared Professional Values** | | | | | | | | | | |
| I1. Respect for data providers, communities, and others' work | | x | (x) | x | | (x) | | (~x) | | (~x) |
| I2. Professionalism: responsibility; competence; and Informed Judgment | (x) | x | | x | x | (x) | | (~x) | (x) | (~x) |
| I3. Truthfulness and integrity: independence; objectivity; transparency | | x | x | x | | | | (~x) | (x) | (~x) |
| **Ethical Principles** | | | | | | | | | | |
| I1. Pursue objectivity by transparent methods and communication | | x | x | x | | | | (~x) | x | (~x) |
| I2. Clarify obligations and roles by area of competency | | x | | | | | | (x) | | (x) |
| I3. Assess Alternatives Impartially | | x | (x) | | | | | | (x) | |
| I4. Avoid Conflicting Interests | | | | | (x) | (x) | | (~x) | | (~x) |
| I5. Avoid Preempted Outcomes | | x | | (x) | | | | (~x) | | (~x) |
| I6. Guard Privileged Information | | | | | | (x) | | (~x) | (x) | (~x) |
| I7. Exhibit Professional Competence | | x | | | | | | | (x) | |
| I8. Maintaining Confidence in Statistics | (x) | (x) | x | (x) | | | | (~x) | x | (~x) |
| I9. Exposing and Reviewing Methods and Findings | | (x) | (x) | (x) | | | | (x) | | x |
| I10. Communicating Ethical Principles | | | | (x) | | | | (x) | x | x |
| I11. Responsibility for the Integrity of the Discipline | | x | (x) | x | | | | | (x) | (~x) |
| I12. Protecting the Interests of Subjects | (x) | | | | | (x) | | (~x) | | (~x) |

# International Guidelines align with ASA Ethical Guidelines

Appendix Table 5d. OECD Good Statistical Practice (2014) and ESSC European Code of Statistical Practice (2017)

| OECD Good Statistical Practice / ESSC European Code of Statistical Practice | O1. A clear legal and institutional framework | O2. Professional independence | O3. Adequacy of resources | O4. Protection of privacy | O5. The right to access administrative sources | O6. Impartiality, objectivity and transparency | O7. Sound methods and professional standards | O8. Quality of statistical outputs and processes | O9. User-friendly access and commitment to respond to user misinterpretation | O10. Coordination of statistical activities | O11. International cooperation | O12. Explore new and alternative data sources and methods |
|---|---|---|---|---|---|---|---|---|---|---|---|---|
| **Institutional Environment** | | | | | | | | | | | | |
| E1. Professional Independence | | x | | | | x | | x | | ~ | ~ | |
| E1bis. Coordination and Cooperation | | (~) | | (~) | | (~) | (~) | (~) | | x | x | |
| E2. Mandate for Data Collection and Access to Data | (x) | | x | x | x | | (x) | | | x | x | |
| E3. Adequacy of Resources | | | x | x | x | | | | | x | x | |
| E4. Commitment to Quality | | x | | x | | x | x | x | x | | | x |
| E5. Statistical Confidentiality and Data Protection | x | x | (x) | x | x | | x | x | | ~ | ~ | |
| E6. Impartiality and Objectivity | | x | | | x | x | x | x | | ~ | ~ | (x) |
| **Statistical Processes** | | | | | | | | | | | | |
| E7. Sound Methodology | | x | | (x) | | x | x | x | x | | (x) (~) | (~) |
| E8. Appropriate Statistical Procedures | | x | | (x) | | x | x | x | x | (~) | (x) (~) | (~) |
| E9. Non-excessive Burden on Respondents | (x) | | | | | | (x) | | (x) | | | x |
| E10. Cost Effectiveness | (x) | (~) | (x) | | | | | (x) | | x | x | |

International Guidelines align with ASA Ethical Guidelines

| OECD Good Statistical Practice<br><br>ESSC European Code of Statistical Practice | O1. A clear legal and institutional framework | O2. Professional independence | O3. Adequacy of resources | O4. Protection of privacy | O5. The right to access administrative sources | O6. Impartiality, objectivity and transparency | O7. Sound methods and professional standards | O8. Quality of statistical outputs and processes | O9. User-friendly access and commitment to respond to user misinterpretation | O10. Coordination of statistical activities | O11. International cooperation | O12. Explore new and alternative data sources and methods |
|---|---|---|---|---|---|---|---|---|---|---|---|---|
| **Statistical Output** | | | | | | | | | | | | |
| E11. Relevance | (x) | (~) | | (~) | | | x | (x) | x | (x) (~) | (x) (~) | (x) (~) |
| E12. Accuracy and Reliability | | x | | | | x | x | x | | (x) (~) | (x) (~) | (x) (~) |
| E13. Timeliness and Punctuality | | (~) | (~) | (~) | | | | (x) | | (x) (~) | (x) (~) | |
| E14. Coherence and Comparability | | x (~) | | | (x) | x (~) | x | x | x | x | (x) | (x) (~) |
| E15. Accessibility and Clarity | (x) | | | | x | x | x | x | x | x (~) | x (~) | |

Appendix Table 5e. OECD Good Statistical Practice (2014) and ISI Statement of Ethics (2023)

| OECD Good Statistical Practice / ISI Statement of Ethics | O1.. A clear legal and institutional framework | O2. Professional independence | O3. Adequacy of resources | O4. Protection of privacy | O5. The right to access administrative sources | O6. Impartiality, objectivity and transparency | O7. Sound methods and professional standards | O8. Quality of statistical outputs and processes | O9. User friendly access and commitment to respond to user misinterpretation | O10. Coordination of statistical activities | O11. International cooperation | O12. Explore new and alternative data sources and methods |
|---|---|---|---|---|---|---|---|---|---|---|---|---|
| **Shared Professional Values** | | | | | | | | | | | | |
| I1. Respect for data providers, communities, and others' work | | | | x | | (x) | | | | (~x) | (~x) | |
| I2. Professionalism: responsibility; competence; and Informed Judgment | | x | (x) | | | x | x | | (x) | (~x) | (~x) | (x) |
| I3. Truthfulness and integrity: independence; objectivity; transparency | | x | | | | x | (x) | | | (~x) | (~x) | (x) |
| **Ethical Principles** | | | | | | | | | | | | |
| I1. Pursue objectivity by transparent methods and communication | | x | (x) | | | x | x | x | x | (~x) | (~x) | (x) |
| I2. Clarify obligations and roles by area of competency | | (x) | | | | | | | | (x) | (x) | |
| I3. Assess Alternatives Impartially | | (x) | (x) | | | x | | | | | | x |
| I4. Avoid Conflicting Interests | | (x) | | | | | | | (x) | (~x) | (~x) | |
| I5. Avoid Preempted Outcomes | | x | | | | (x) | | | | (~x) | (~x) | (x) |

International Guidelines align with ASA Ethical Guidelines

| OECD Good Statistical Practice / ISI Statement of Ethics | O1.. A clear legal and institutional framework | O2. Professional independence | O3. Adequacy of resources | O4. Protection of privacy | O5. The right to access administrative sources | O6. Impartiality, objectivity and transparency | O7. Sound methods and professional standards | O8. Quality of statistical outputs and processes | O9. User friendly access and commitment to respond to user misinterpretation | O10. Coordination of statistical activities | O11. International cooperation | O12. Explore new and alternative data sources and methods |
|---|---|---|---|---|---|---|---|---|---|---|---|---|
| I6. Guard Privileged Information | | | | x | | | | | | (~x) | (~x) | |
| I7. Exhibit Professional Competence | | | | | | | (x) | | | | | x |
| I8. Maintaining Confidence in Statistics | | (x) | (x) | | | (x) | | x | (x) | (~x) | (~x) | x |
| I9. Exposing and Reviewing Methods and Findings | | (x) | | | | | | | (x) | (x) | x | x |
| I10. Communicating Ethical Principles | | (x) | | | | | | | | (x) | x | |
| I11. Responsibility for the Integrity of the Discipline | | (x) | | | | x | | | x | | (~x) | (x) |
| I12. Protecting the Interests of Subjects | | | | (x) | | | | | | (~x) | (~x) | |

Appendix Table 5f. ESSC European Code of Statistical Practice (2017) and ISI Statement of Ethics (2023)

| ESSC European Code of Practice / ISI Statement of Ethics | E1. Professional Independence | E1bis. Coordination and Cooperation | E2. Mandate for Data Collection and Access to Data | E3. Adequacy of Resources | E4. Commitment to Quality | E5. Statistical Confidentiality and Data Protection | E6. Impartiality and Objectivity | E7. Sound Methodology | E8. Appropriate Statistical Procedures | E9. Non-excessive Burden on Respondents | E10. Cost Effectiveness | E11. Relevance | E12. Accuracy and Reliability | E13. Timeliness and Punctuality | E14. Coherence and Comparability | E15. Accessibility and Clarity |
|---|---|---|---|---|---|---|---|---|---|---|---|---|---|---|---|---|
| **Shared Professional Values** | | | | | | | | | | | | | | | | |
| I1. Respect for data providers, communities, and others' work | | (~x) | (x) | | | x | | | | (x) | | | | | | |
| I2. Professionalism: responsibility; competence; and Informed Judgment | (x) | x, (~x) | (x) | (x) | x | | x | x | x | | | x, (~x) | (x) | x, (~x) | | (x) |
| I3. Truthfulness and integrity: independence; objectivity; transparency | x | x,(~x) | (x), (~x) | | x | | x | | | | | | | x | x, (~x) | x |
| **Ethical Principles** | | | | | | | | | | | | | | | | |
| I1. Pursue objectivity by transparent methods and communication | x | x, (~x) | (x), (~x) | | x | | x | x | x | | | | | x | x, (~x) | x |
| I2. Clarify obligations and roles by area of competency | | x | x | | | | | | | | | | | | (x) | |
| I3. Assess Alternatives Impartially | x | x | | (x) | x | | x | x | x | (x) | (x) | | | | | x |
| I4. Avoid Conflicting Interests | (x) | (~x) | (x), (~x) | | | | | | | | | | | | | |
| I5. Avoid Preempted Outcomes | x | x, (~x) | (x), (~x) | | x | | x | x | x | | | | x, (~x) | | | |
| I6. Guard Privileged Information | | (~x) | | | | x | | | | | | | | | | |

International Guidelines align with ASA Ethical Guidelines

| ESSC European Code of Practice / ISI Statement of Ethics | E1. Professional Independence | E1bis. Coordination and Cooperation | E2. Mandate for Data Collection and Access to Data | E3. Adequacy of Resources | E4. Commitment to Quality | E5. Statistical Confidentiality and Data Protection | E6. Impartiality and Objectivity | E7. Sound Methodology | E8. Appropriate Statistical Procedures | E9. Non-excessive Burden on Respondents | E10. Cost Effectiveness | E11. Relevance | E12. Accuracy and Reliability | E13. Timeliness and Punctuality | E14. Coherence and Comparability | E15. Accessibility and Clarity |
|---|---|---|---|---|---|---|---|---|---|---|---|---|---|---|---|---|
| I7. Exhibit Professional Competence | | x | | | x | | | | | | | | | | | |
| I8. Maintaining Confidence in Statistics | (x) | (x), (~x) | (x) | | x | | x | | | | | | x | x, (~x) | | |
| I9. Exposing and Reviewing Methods and Findings | (x) | (x),(~x) | | | x | | x | | | | | | | x, (~x) | | x |
| I10. Communicating Ethical Principles | | x | | | x | (x) | | | | | | | | | (x) | |
| I11. Responsibility for the Integrity of the Discipline | | x, (~x) | (x), (~x) | | x | | | | | | | | x | | | |
| I12. Protecting the Interests of Subjects | (x) | x, (~x) | | | | x | | | | (x) | | (~x) | | (~x) | | |